\documentclass[pdflatex, sn-mathphys-num]{sn-jnl}

\usepackage{graphicx}%
\usepackage{multirow}%
\usepackage{amsmath,amssymb,amsfonts}%
\usepackage{amsthm}%
\usepackage{mathrsfs}%
\usepackage[title]{appendix}%
\usepackage{xcolor}%
\usepackage{textcomp}%
\usepackage{manyfoot}%
\usepackage{booktabs}%
\usepackage{algorithm}%
\usepackage{algorithmicx}%
\usepackage{algpseudocode}%
\usepackage{listings}%
\usepackage{braket}
\usepackage{url}
\usepackage{float}
\usepackage{lineno}
\usepackage[normalem]{ulem} 

\theoremstyle{thmstyleone}%
\theoremstyle{thmstyletwo}%

\theoremstyle{thmstylethree}%

\begin{document}

\title[Article Title]{Quantum Generative Modeling of Single-Cell Transcriptomes: Capturing Gene-Gene and Cell-Cell Interactions}

\author*[1,2,3]{\fnm{Selim} \sur{Romero}}\email{ssromerogon@tamu.edu}

\author[1]{\fnm{Vignesh} \sur{S Kumar}}\email{vignesh\_sk@tamu.edu}

\author[2,3]{\fnm{Robert S.} \sur{Chapkin}}\email{r-chapkin@tamu.edu}

\author*[1,3,4]{\fnm{James J.} \sur{Cai}}\email{jcai@tamu.edu}

\affil[1]{\orgdiv{Department of Veterinary Integrative Biosciences}, \orgname{Texas A\&M University} 
}

\affil[2]{\orgdiv{Department of Nutrition}, \orgname{Texas A\&M University}
}

\affil[3]{\orgdiv{CPRIT Single Cell Data Science Core}, \orgname{Texas A\&M University}
}

\affil[4]{\orgdiv{Department of Biochemistry \& Molecular Biology}, \orgname{University of Georgia}
}

\abstract{
Single-cell RNA sequencing (scRNA-seq) data simulation is limited by classical methods that primarily rely on linear correlations, failing to capture the intrinsic, nonlinear dependencies. Existing simulators do not jointly model gene-gene regulatory interactions and cell-cell communication. We introduce qSimCells, a novel quantum computing-based simulator that employs entanglement to model intra- and inter-cellular interactions, generating realistic single-cell transcriptomic data from heterogeneous cell populations. The core innovation is a quantum kernel that uses a parameterized quantum circuit with CNOT gates to encode complex, nonlinear gene regulatory networks (GRNs) together with cell-cell communication topologies. 
 By explicitly programming the entanglement architecture, the simulator establishes a known generative ground truth for both regulatory and communication pathways.
The resulting synthetic data exhibits dependencies arising from the joint probability structure of the quantum circuit. Notably, standard correlation-based analyses (Pearson and Spearman) fail to recover the programmed causal relationships and instead report spurious associations driven by high baseline gene-expression probabilities. Furthermore, applying cell-cell communication detection to the simulated data serves as an internal consistency check, where CellChat, a widely adopted communication detection framework, correctly identifies the true ligand-receptor pairs when inter-state entanglement is active, revealing a robust, up to ~98-fold relative increase in inferred communication probability. These results demonstrate that the quantum kernel is instrumental for producing high-fidelity benchmark datasets 
with known ground truth. More broadly, they highlight the limitations of conventional correlation-based inference methods and underscore the need for advanced analytical approaches capable of capturing the complex structural dependencies underlying gene regulation and cell–cell communication.}

%
%
%


\maketitle
\section{Introduction}\label{sec_intro}
Single-cell RNA sequencing (scRNA-seq) has transformed modern biology by enabling gene expression profiling at single-cell resolutions \cite{zheng2017massively,haque2017practical}. This capability allows researchers to explore the molecular signatures that define individual cell identities and functions, thereby revealing the complexity of cellular heterogeneity. Gene expression, however, is not a linear process--it arises from intricate, nonlinear interactions among genes that form dynamic gene regulatory networks (GRNs) essential for cellular functions \cite{GRN_ref}.

Simulating single-cell data is instrumental in developing and benchmarking computational approaches for understanding cellular heterogeneity. Simulators produce synthetic datasets with a known ground truth, enabling rigorous evaluation of GRN construction, clustering and cell-cell communication inference. Without a simulator that faithfully encodes causal structure in a GRN, for example, it is impossible to assess whether an inference algorithm recovers true regulatory relationships or merely statistical artifacts. Existing tools for simulating single-cell data are far from scarce. However, generating realistic single-cell simulations remains challenging, as it requires capturing not only the intricate gene-gene regulatory dynamics within cells but also the ligand-receptor signaling that coordinates behavior across cells \cite{era,sergio,multisim}. 

Indeed, recent benchmark studies highlight that classical simulators struggle to accommodate complex designs and yield unreliable performance estimates, emphasizing the need for more expressive models \cite{shaky}. To address this gap, quantum generative modeling (QGM) is a plausible  approach to achieve the high expressivity required for complex probability distributions, offering an advanced approach to synthetic data generation \cite{kyriienko2024,gao2022enhancing}.

Most available simulators prioritize intracellular regulation, while models of cell-cell communication remain immature. 
SERGIO~\cite{sergio} models gene regulatory networks via Langevin stochastic differential equations within a single cell type but has no intercellular coupling; scMultiSim~\cite{multisim} advances this with CIF (Cell Identity Factor)-based Beta-Poisson kinetics across a lineage tree and an optional proximity-dependent ligand--receptor signalling layer, but treats intercellular effects as an additive module separate from the core GRN. Neither encodes a direct, gene-level regulatory cascade spanning multiple cell types in a unified generative model.
Meanwhile, attempts to model cell-cell communication remain rudimentary, typically approximating ligand–receptor signaling through static mappings with fixed communication probabilities \cite{comparison, cesaro2025advances}. In a typical setup, to make two cells “interact”, the simulator goes through each sender cell and finds designated receiver cells within a neighborhood \cite{armingol2021deciphering,cellchat}. For every pair of sender-receiver cells, it checks a list of matching ligand-receptor genes. Then, for each match, it slightly increases the receptor cell’s gene expression based on how strongly the sender cell expresses the ligand through linear correlation. This exhaustive, pairwise procedure must be repeated for every cell and gene pair, is slow, cumbersome, and provides no mechanistic intercellular feedback. 

There is currently no platform capable of simulating both intracellular and intercellular dynamics in an integrated manner. We previously developed the quantum single-cell GRN framework (qscGRN), using a parameterized quantum circuit to infer GRNs from single-cell data \cite{qgrn}. By leveraging qubit entanglement to represent gene-gene dependencies and optimizing a Kullback-Leibler divergence-based loss function within a hybrid quantum-classical loop, qscGRN demonstrated the potential of quantum computing to uncover complex gene regulatory relationships beyond the reach of conventional statistical models \cite{qgrn}, part of a growing body of quantum approaches to biological data analysis including tensor-network GRN inference for single-cell data \cite{larrarte2025tensor}, quantum computing for single-cell omics and cell-based therapeutics \cite{bose2026advancing}, and empirical quantum kernel benchmarking on gene expression data \cite{ghosh2025empirical}. Building upon this foundation, the present work introduces a hybrid quantum-classical simulator that extends quantum principles--specifically \textit{superposition} and \textit{entanglement}--to model single-cell gene expression and intercellular communication. In our design, qubits serve as analogues for genes or molecular features. Custom rotation gates initialize each qubit to represent basal gene expression levels, while CNOT gates introduce entangled relationships between qubits, thereby encoding the nonlinear topology of GRNs. This enables the generation of diverse and biologically realistic gene expression patterns with explicit causal structure not naturally expressible in classical correlation-based simulators. Furthermore, by entangling ligand and receptor gene qubits, our framework directly simulates cell–cell communication, capturing molecular crosstalk between distinct cell types. The key contribution of this study is a quantum-based simulation framework that models both intra-cellular regulatory mechanisms and inter-cellular signaling. By explicitly representing ligand-receptor interactions across different cell types, the proposed method provides a more direct and mechanistic view of cell-cell communication than classical neighborhood-based approaches.

\section{Methods}\label{sec_methods}
\begin{figure}[h!] 
    \centering 
    \includegraphics[width=1.0\textwidth]{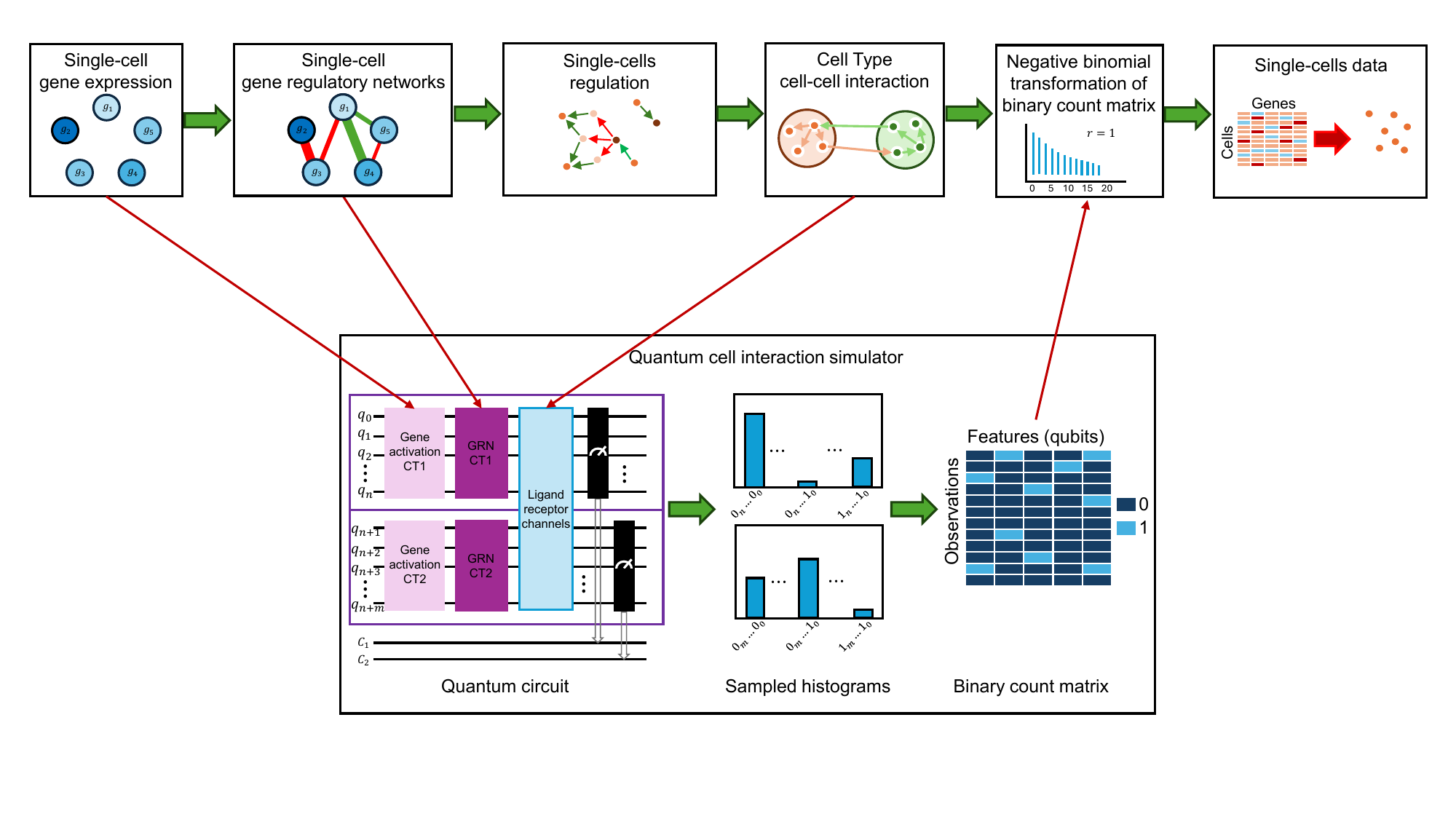}
    \caption{\textbf{Quantum-simulated single-cell data framework.} Our framework utilizes a quantum kernel, divided into three key sections, to generate realistic single-cell data. First, independent feature (gene) activation is simulated for two distinct cell types (CT1, cell type 1, and CT2, cell type 2). Second, a gene regu- latory network (GRN) is established using controlled-NOT (CX) gates between source and target qubits, modeling gene-gene interactions. Third, inter-cellular (inter-state) communication channels are introduced and enhanced by CX gates, representing interactions between cell types. The simulation generates sampled binary histograms for CT1 and CT2, encoding complex inter- and intra-cellular communication within the feature (qubit) states. These histograms are then converted into a binary matrix, where each entry represents the simultaneous activation of features (genes) within individual simulated cells. The final step involves transforming this binary matrix into gene expression profile using a negative binomial augmentation, thus creating a more biologically realistic synthetic single-cell data.
    } 
    \label{fig:scheme}
\end{figure}

We introduce qSimCells, as illustrated in Fig. \ref{fig:scheme}, a quantum computing-based generative framework for single-cell data simulation.

The core of framework is a quantum kernel, designed to capture complex, entanglement-enforced  dependencies through quantum entanglement. 
This kernel enables the simulation of both intra-cellular interactions within a quantum-entangled GRN and inter-cellular communication via ligand–receptor (LR) pair entanglement across distinct statevectors \cite{qgrn}. 

Quantum computing offers a very interesting perspective for modeling relationships between genes as well as between cells of different types \cite{qgrn}. Our proposition utilizes a Parameterized Quantum Circuit (PQC) where the initial cell state $\ket{\psi_0}$ is prepared using user-defined gene activation angles $\theta_i$.

The state initialization involves applying $Y$-axis rotation ($R_y$) to each of the $n$ qubits, which starts in the ground state $\ket{0}$:
\begin{equation}\label{eq1:ang_act}
    \ket{\psi_0} =  \otimes_{i=0}^{n-1} R_y(\theta_i) \ket{0_i}.
\end{equation}
Next, to integrate GRN interactions, we couple the expression of one gene to another using a Controlled-NOT ($CX$) gate. This process encodes a directed causal dependency: the targeted gene's activation is conditioned on the control gene's state, establishing explicit time-ordered regulatory relationships \cite{qgrn}. While the present $Ry+CX$ circuit is classically simulable, the quantum framework provides an architectural advantage as a natural interface for encoding such dependencies.

The resulting entangled state $\ket{\psi_1}$ is obtained by applying a sequence of $M$ $CX$ gates, where the sequence is crucial due to the non-commuting nature of the gates, potentially mimicking cascade activations:
\begin{equation}\label{eq2:grn_entangle}
    \ket{\psi_1} = \left( \prod_{k=0}^{M-1} CX_{c_k,t_k} \right)_{\text{time-ordered}} \ket{\psi_0}.
\end{equation}

The time ordered product means that the gates are applied sequentially, with $CX_{c_1,t_1}$ applied first and $CX_{c_{M-1},t_{M-1}}$ applied last. The sequence of $CX$ gates is explicitly defined by the list of control-target pairs $L=\{(c_k,t_k)\}_{k=0}^{M-1}$, which represents the GRN topology \cite{qgrn}. This ordering is significant because $CX$ gates sharing a qubit do not commute in general, meaning that swapping two such gates can change the resulting quantum state. In practice, the implementation uses the $CRX(\pi)$ gate, which is equivalent to $CX$ up to a global phase factor and leaves all measurement outcome probabilities unchanged; however, $CRX(\lambda)$ with $\lambda \in (0, \pi)$ enables users to tune the strength of partial entanglement between gene pairs. Throughout this work we retain the $CX$ notation for clarity, with the understanding that the full rotation angle $\lambda = \pi$ is used, unless specified a different angle.

A key advantage of quantum computing lies in its ability to combine states via the tensor product, which drastically expands the Hilbert space and enables the representation of  complex interactions \cite{nielsen}. 
More importantly, to simulate two distinct cell types, we can prepare two independent cell states, $\ket{\psi_1}$  (on $n$ qubits) and $\ket{\psi_1'}$ (on $n'$ qubits), which are initially defined on disjoint sets of qubits. The combined state is formed by their tensor product:
\begin{equation}\label{eq3:augment}
    \ket{\psi_2} = \ket{\psi_1} \otimes \ket{\psi_1'}.
\end{equation}
The resulting composite state $\ket{\psi_2}$ lives in a Hilbert space of dimensions $2^{n+n'}$ and serves as the foundation for modeling inter-state interactions (e.g., between different cell types) by applying subsequent entangling gates across the $n$ and $n'$ qubit registers, similar to the coupling defined in Eq. \ref{eq2:grn_entangle} \cite{nielsen}. 
Under the assumption that there is no interaction between two cell types (a baseline model), we keep the measurements for the two different cell state registers separated, allowing for independent analysis.

\subsection{Final entanglement and quantum simulation}\label{subsec_simulation}
To complete the model, we introduce cell-cell interactions (between two cell types) by applying a final set of entangling gates. Those gates are put across the combined $\ket{\psi_2}$ state to establish inter-state interactions (e.g., cell-to-cell communication between the $n$ and $n'$ gene registers). This is achieved by applying a fixed sequence of $K$ $CX$ gates. The resulting final state, $\ket{\psi_{\text{final}}}$, is defined as:

\begin{equation}\label{eq4:inter_entangle}
\ket{\psi_{\text{final}}} = \left( \prod_{k=0}^{K-1}CX_{c'_k,t'_k} \right)_{\text{time-ordered}} \ket{\psi_2}.
\end{equation}

The indices $c_k'$ and $t_k'$ here represent global qubit indices that span both the $n$ and $n'$ registers. The $\ket{\psi_{\text{final}}}$ state is obtained and executed using the Qiskit quantum computing framework developed by IBM \cite{qiskit}. Our methodology supports two implementations: for model prototyping, the circuit is run on the local quantum computer simulator, the AerSimulator; for realistic results incorporating quantum hardware noise, the circuit is executed on an IBM Quantum computer \cite{qiskit}.
In both cases, the probability distribution of the final state is sampled by executing a fixed number of measurement shots using the SamplerV2 primitive. The number of shots, $N_{\text{shots}}$ is set equal to the total number of simulated single-cell observations $m$. All stochastic steps are seeded for reproducibility (seed = 42 throughout).
Furthermore, the raw output bit strings from the simulation are reversed (e.g., $b=b_{n-1}...b_0$)  to align with the logical gene indices $i=0 \cdots n-1$, compensating for the little-endian ordering convention of the quantum simulator.

\subsection{scRNA-seq count matrix generation}\label{subsec_count_mat}
The simulation process begins by measuring the final quantum state $\ket{\psi_{\text{final}}}$ multiple times to obtain a histogram of the measurement outcomes. These outcomes are recorded into two distinct classical registers, allowing the probability distribution (marginalized for each cell state) to be assessed separately. This distribution is then used to assess the co-occurrence of gene activation (features) across $m$ observations (simulated single-cells) from the $N_{shots}$ quantum circuit events.
A binary count matrix $X' \in \mathbb{R}^{m \times n}$ is first constructed from this measurement histogram. Each measured bit string, $b=b_0b_1\ldots b_{n-1}$, corresponds to a single-cell observation where the value $b_j \in \{0,1\}$ indicates the deactivation or activation 
(expression) of gene $j$, respectively. If a bit string $b$ has a measured count of $C(b)$, it contributes $C(b)$ rows to $X'$, where each row represents a simulated cell and each column represents a gene.

\begin{equation}
    X'_{ij} = 
    \begin{cases} 
        1 & \text{if gene } j \text{ is `ON' for cell } i \\
        0 & \text{if gene } j \text{ is `OFF' for cell } i
    \end{cases}
\end{equation}

The generated binary count matrix $X'$ is then transformed to incorporate the continuous and noisy characteristics of gene expression counts observed in real single-cell data. This is achieved by multiplying the observed `ON' states ($X_{ij}'=1$) by a value sampled from the Negative Binomial distribution, a function commonly attributed to the overdispersed count data characteristic of scRNA-seq \cite {GNB}. The final gene count matrix $X \in \mathbb{R}^{m \times n}$ is calculated as:

\begin{equation}
    X_{ij} = NB(r_j,p_j) X_{ij}'.
\end{equation}

Here, $NB(r_j,p_j)$ represents a random variate (or single random sample) drawn from the Negative Binomial (NB) distribution parameterized by gene-specific parameters, where $r_j$ (often related to dispersion) is the number of successful trials, and $p_j$ is the probability of success. $p_j$ is defined as $p_j=r_j/(\mu_j+r_j)$, where $\mu_j$ is the $j$-th gene mean. This transformation ensures that the final count $X_{ij}$ remains 0 if the gene was not activated in the quantum measurement ($X_{ij}'=0$), but if $X_{ij}'=1$, the expression level follows the stochastic, overdispersed behavior of a real gene. The parameters $r_j$ and $p_j$ are typically designed to mimic real scRNA-seq data to match the marginal statistics of the genes being modeled \cite{splatter,shaky}.

\subsection{Inferring gene regulatory networks with simulated data}\label{sec:network}
To benchmark the complexity and entanglement-structured nature of synthetic data, we applied classical GRN inference methods \cite{ali2025review} to the qSimCells simulated data.
Prior to inference, the raw scRNA-seq count matrix $X$ was preprocessed following standard single-cell analysis practices \cite{sajita,luecken}:
\begin{enumerate}
    \item \textbf{Normalization:} Total counts were normalized to $10,000$ per cell to correct for sequencing depth differences.
    \item \textbf{Transformation:} The data was $\log_{1p}$-transformed to stabilize the variance and mitigate the influence of large count magnitudes \cite{bacher}.
    \item \textbf{Scaling:} The data was standardized (Z-score scaled) per gene to ensure all features contributed equally to the correlation metrics.
\end{enumerate}
Using this preprocessed matrix, gene-gene correlation matrices were computed across the cells using both the Pearson (linear) and Spearman (monotonic non-linear) correlation coefficients \cite{faith2007large}. Advanced tree-based GRN inference methods such as GENIE3 \cite{huynh2010inferring} and GRNBoost2 \cite{moerman2019grnboost2} were evaluated in the 4-method benchmark described below (Fig.~\ref{fig:grn_corr}).
An adjacency matrix was then constructed by retaining only edges where the absolute correlation value exceeded the threshold ($|\text{Corr}| > \tau$). The threshold was varied across the range $[0.3, 0.7]$ and the qualitative result for Pearson and Spearman, failure to recover the programmed $CX$ topology, remained consistent across all tested values, confirming that the finding is not an artifact of the specific threshold chosen. Network graphs can be explored using the \text{NetworkX} package \cite{hagberg}; however, for the benchmark reported here, only the thresholded adjacency matrix was used to compute edge-recovery frequency across the 10 independently seeded replicates and 5 thresholds (Fig.~\ref{fig:grn_corr}).

\section{Results}\label{sec_res}
Our initial simulation demonstrates the capability of the proposed quantum kernel to model both intra-cellular regulations within (GRNs) and inter-cellular communication between distinct cell states. In this proof-of-concept study, we simulated a system consisting of five genes ($n=5$) for Cell Type 1 (CT1) and five genes ($n'=5$) for Cell Type 2 (CT2) allowing us to evaluate the framework's ability to capture gene-gene dependencies and LR-mediated cross-talk between cells.

\subsection{Parameter initialization and qubit mapping}\label{subsec:param_init}
The initial self-activation level for each gene is set by its corresponding rotation angle $\theta_i$, as defined in Eq.~\ref{eq1:ang_act}. The normalized rotation parameter $a_i = \theta_i/\pi$ serves as the design input for the $R_y$ initial activation operation, while the exact activation probability is $P_i = \sin^2(\theta_i/2)$. Both are listed in Table~\ref{tab:initial_angles}.

To establish an unambiguous indexing system, the system's $n+n'=10$ qubits are mapped sequentially, following the augmentation in Eq. \ref{eq3:augment}. This creates a global gene index $g_i$ (where $i=0$ to $9$) that is identical to the qubit index $q_i$. Specifically:
\begin{itemize}
    \item Genes $g_0$ to $g_4$ correspond to CT1.
    \item Genes $g_5$ to $g_9$ correspond to CT2.
\end{itemize}
We interchangeably utilize the global index notation $q_i \leftrightarrow g_i$ throughout the remainder of this work. By construction, each register is entirely absent from the other cell type's circuit: CT1 qubits ($g_0$--$g_4$) produce no expression of CT2 genes ($g_5$--$g_9$), and vice versa. This structural zero-expression is intentional, ensuring that any cross-type signal detected in the interacting condition arises exclusively from the inter-register $CX$ entanglement gates, not from baseline co-expression.

\begin{table}[h!]
 \centering
 \caption{Initial gene activation parameters and qubit mapping}
 \label{tab:initial_angles}
 \begin{tabular}{ccccc}
 \toprule
 \textbf{Global Index ($q_i/g_i$)} & \textbf{$a_i = \theta_i/\pi$} & \textbf{$P_i = \sin^2(\theta_i/2)$} & \textbf{Cell Type} & \textbf{Local Gene Index}\\
 \midrule
 $q_0 \equiv g_0$ & $0.2$ & $0.095$ & CT1 & $g_0^{\text{CT1}}$ \\
 $q_1 \equiv g_1$ & $0.1$ & $0.024$ & CT1 & $g_1^{\text{CT1}}$ \\
 $q_2 \equiv g_2$ & $0.4$ & $0.345$ & CT1 & $g_2^{\text{CT1}}$ \\
 $q_3 \equiv g_3$ & $0.9$ & $0.976$ & CT1 & $g_3^{\text{CT1}}$ \\
 $q_4 \equiv g_4$ & $0.8$ & $0.905$ & CT1 & $g_4^{\text{CT1}}$ \\
 \midrule
 $q_5 \equiv g_5$ & $0.2$ & $0.095$ & CT2 & $g_0^{\text{CT2}}$ \\
 $q_6 \equiv g_6$ & $0.3$ & $0.206$ & CT2 & $g_1^{\text{CT2}}$ \\
 $q_7 \equiv g_7$ & $0.2$ & $0.095$ & CT2 & $g_2^{\text{CT2}}$ \\
 $q_8 \equiv g_8$ & $0.7$ & $0.794$ & CT2 & $g_3^{\text{CT2}}$ \\
 $q_9 \equiv g_9$ & $0.5$ & $0.500$ & CT2 & $g_4^{\text{CT2}}$ \\
 \bottomrule
 \end{tabular}
\end{table}

\subsection{Entanglement and cascade activation}
The entanglement topology governing both intra- and inter-state gene regulatory interactions was defined by the control-target list $L$. In this framework, intra-state interactions represent gene-gene regulation within a single cell type, while inter-state interactions represent ligand-receptor communication between two distinct cell types.

\subsubsection{Case 1: Inter-state cascade}
In the first case, we designed a specific inter-state cascade utilizing the entanglement topology $L_1 = \{ (3, 5), (5, 7), (7, 0) \}$. This configuration, applied in the $\ket{\psi_{\text{final}}}$ stage (Eq. \ref{eq4:inter_entangle}), models a multi-step regulatory path spanning both cell types:

\begin{quote}
\begin{itemize}
    \item \textbf{Step 1 (CT1 $\to$ CT2):} Gene $g_3^{\text{CT1}}$ (qubit $q_3$) activates gene $g_0^{\text{CT2}}$ (qubit $q_5$).
    \item \textbf{Step 2 (CT2 $\to$ CT2):} Gene $g_0^{\text{CT2}}$ (qubit $q_5$) activates gene $g_2^{\text{CT2}}$ (qubit $q_7$).
    \item \textbf{Step 3 (CT2 $\to$ CT1):} Gene $g_2^{\text{CT2}}$ (qubit $q_7$) activates gene $g_0^{\text{CT1}}$ (qubit $q_0$).
\end{itemize}
\end{quote}

The resulting entangled state $\ket{\psi_{\text{final}}}$ reflects this cascade. The entangling topology and the resulting measurement histogram are shown in Fig. \ref{fig:sampling_qc}A.

\subsubsection{Case 2: Non-interacting control}
To highlight the effect of the inter-state communication, we performed a control experiment. We used the same activation angles (Table \ref{tab:initial_angles}) but replaced the complex cascade with a simple, non-communicating intra-state entangler $L_2 = \{ (2, 1) \}$. This design models an isolated regulation within $\text{CT1}$ (affecting $g_2^{\text{CT1}}$ and $g_1^{\text{CT1}}$) and enforces no communication between $\text{CT1}$ and $\text{CT2}$.

\begin{figure}[h!]
 \centering
 \includegraphics[width=1.0\textwidth]{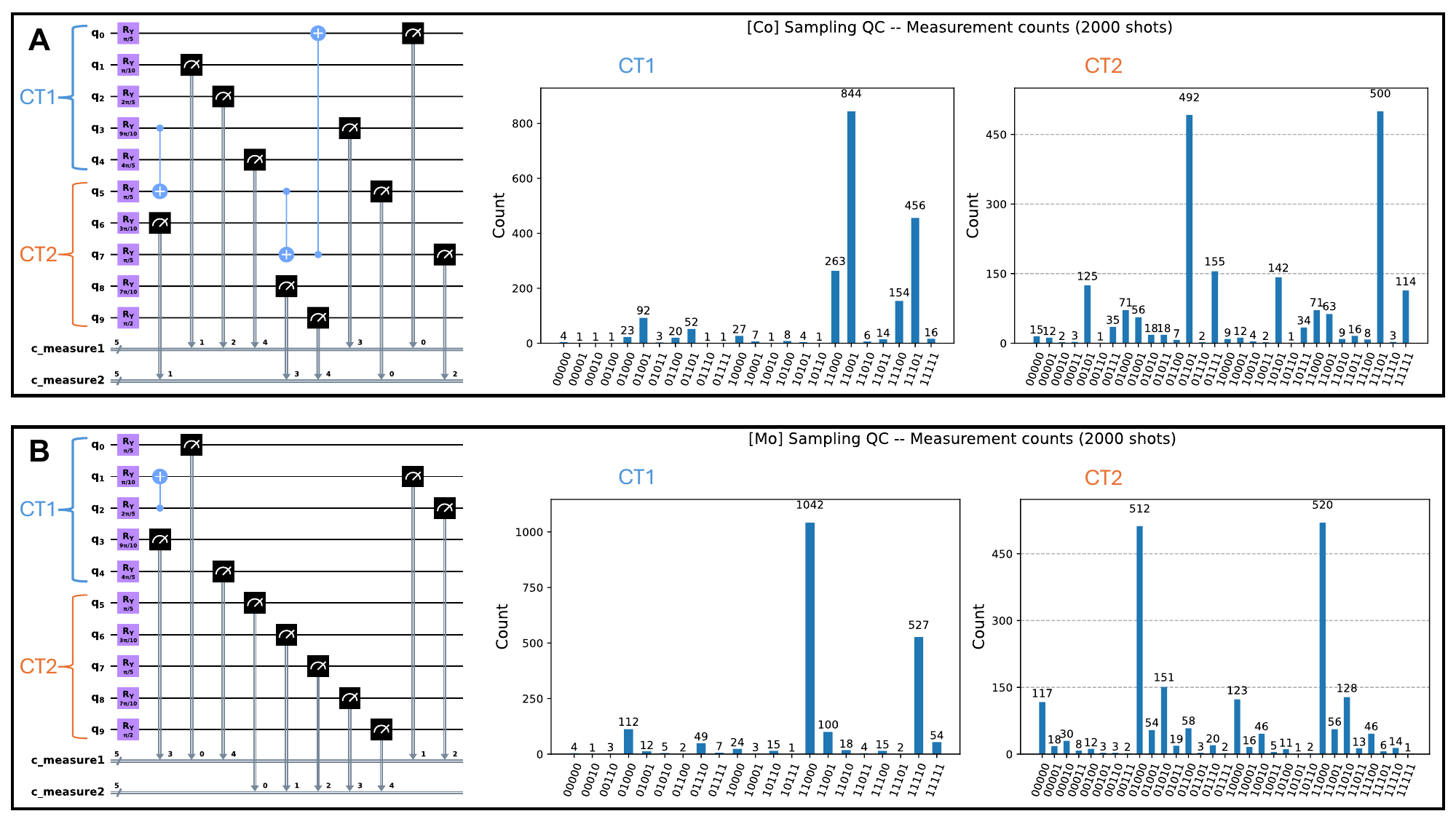}
 \caption{\textbf{Quantum circuit sampling.} A) shows inter-state interactions modeled by $L_1 = \{ (3, 5), (5, 7), (7, 0) \}$ entanglement interactions on
 the quantum circuit, and the corresponding measurements per $\ket{\psi_0}$  state. B) shows non-communicating intra-state modeled by $L_2 = \{ (2, 1) \}$
 entanglement on the quantum circuit, and the corresponding measurements per  $\ket{\psi_0}$ state. Shots are set equal to the number of simulated cells ($N_{\text{shots}}=m=2{,}000$); re-simulation at $N_{\text{shots}}=100{,}000$  confirms that the relative ranking of outcomes is preserved and that only the direct $CX$ targets ($q_5$, $q_7$, $q_0$) show the probability boost consistent with the  cascade activation product (see Supplemental Information Fig. S1.)}
 \label{fig:sampling_qc}
\end{figure}

Fig. \ref{fig:sampling_qc}B shows the resulting measurement histogram for this non-cascading topology. A visual comparison between the histograms in Fig. \ref{fig:sampling_qc}A and $\ref{fig:sampling_qc}\text{B}$ demonstrates how the programmed entangling topology significantly alters the final co-expression patterns.

\subsection{Negative binomial augmentation for synthetic scRNA-seq data}\label{subsec_count_mat}

Following the quantum simulation and measurement, the resulting $X'$ matrix (the binary count matrix) is utilized to generate the final scRNA-seq count data, as described in Section \ref{subsec_count_mat} \cite{GNB}. This step involves transforming the binary gene activation states into continuous count data by sampling from the NB distribution to accurately model the biological noise and overdispersion characteristic of single-cell sequencing \cite{GNB}.

For demonstration and simplified analysis, the gene-specific mean ($\mu_i$) and dispersion ($r_i$) parameters were uniformly set to $\mu_i=5$ and $r_i=1$ across all model genes in both cell types (CT1 and CT2). This modular approach ensures that the initial regulatory pattern is governed by the rotation angle $\theta_i$ and the GRN entanglement topology, while the final expression level (magnitude) and stochasticity are independently controlled by the customizable NB parameters $\mu_i$ and $r_i$.

To stabilize the cell type representations and provide a robust baseline for expression, we augmented the gene set by including 50 Housekeeping Genes (HKGs) \cite{housekeep}. These were assigned high, stable expression parameters: $\mu_{\text{HKG}} = 80$ and $r_{\text{HKG}}=6$.

\subsection{Effect of including inter-state interaction on simulated cell populations}
\begin{figure}[h!]
 \centering 
 \includegraphics[width=1.0\textwidth]{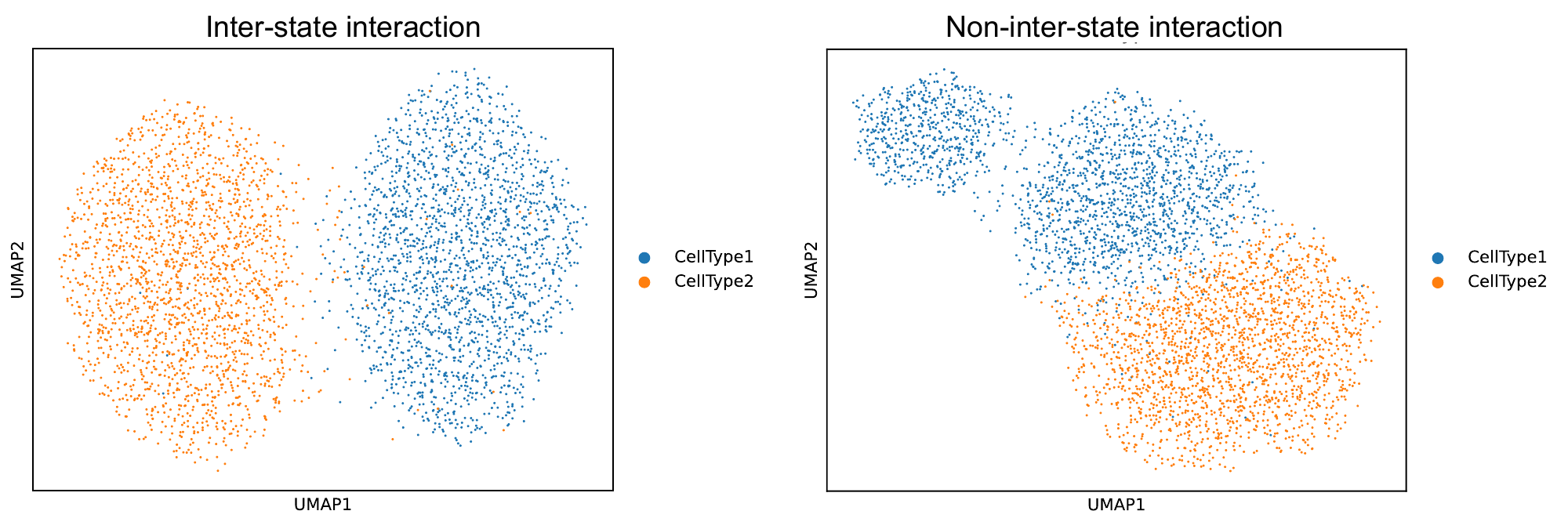}
 \caption{\textbf{UMAP visualization of simulated scRNA-seq data.} The UMAP plot displays the synthetic single-cell data generated under two different entanglement scenarios (Case 1: Inter-state cascade; Case 2: Non-interacting control), showing the resulting lineage separation between CT1 and CT2 populations.}
 \label{fig:scRNA_seq_synthetic}
\end{figure}

The synthetic scRNA-seq data, visualized in the UMAP (Uniform Manifold Approximation and Projection) plot in Fig. \ref{fig:scRNA_seq_synthetic}, reveals a clear distinction between cell populations simulated with the two different entanglement scenarios: with and without inter-state interaction.

\begin{itemize}
    \item \textbf{Inter-state cascade (case 1):} The data generated with the inter-state regulatory cascade shows a distinct separation of the CT1 and CT2 lineages in Fig. \ref{fig:scRNA_seq_synthetic}. This pronounced separation is due to two factors: the enforced cross-type expression enhancement, and the structural exclusion of gene expression, where CT2 genes ($q_5$–$q_9$) have zero expression in CT1 cells and, conversely, CT1 genes ($q_0$–$q_4$) have zero expression in CT2 cells \cite{inference}. This structural sparsity, combined with entanglement, drives the populations into structurally different high-dimensional states.
    \item \textbf{Non-interacting control (case 2):} Conversely, the non-interacting control experiment (where entanglement was restricted to only $L_2 = \{ (2, 1) \}$) exhibits a more mixed or less pronounced separation. In this case, the expression pattern is primarily dominated by the uniformly highly expressed HKGs and the non-interacted model genes. The inter-state cascade is essential for providing the unique, non-linear expression patterns that maximize the separation of the cell states \cite{inference}.
\end{itemize}
The observed lineage separation confirms that quantum entanglement, specifically when programmed to facilitate cross-state regulation, is the dominant factor shaping the distinct expression profiles of CT1 and CT2.

\subsection{Synthetic scRNA-seq data analysis and classical predictions}
The synthetic scRNA-seq data produced from quantum computing kernel has complex relationships that would be hardly embedded from classical regime, but we can still see if from classical regime, we could get the grasp of what was embedded. To this purpose, we propose two analyses, one being the GRN benchmark evaluating four classical inference methods (Pearson, Spearman, GRNBoost2, and GENIE3). While the second is to apply CellChat to see if we can get the cell communication difference between the two previous cases.

\subsubsection{Inferred gene regulatory networks from synthetic data}

\begin{figure}[H]
  \centering
  \includegraphics[width=1.0\textwidth]{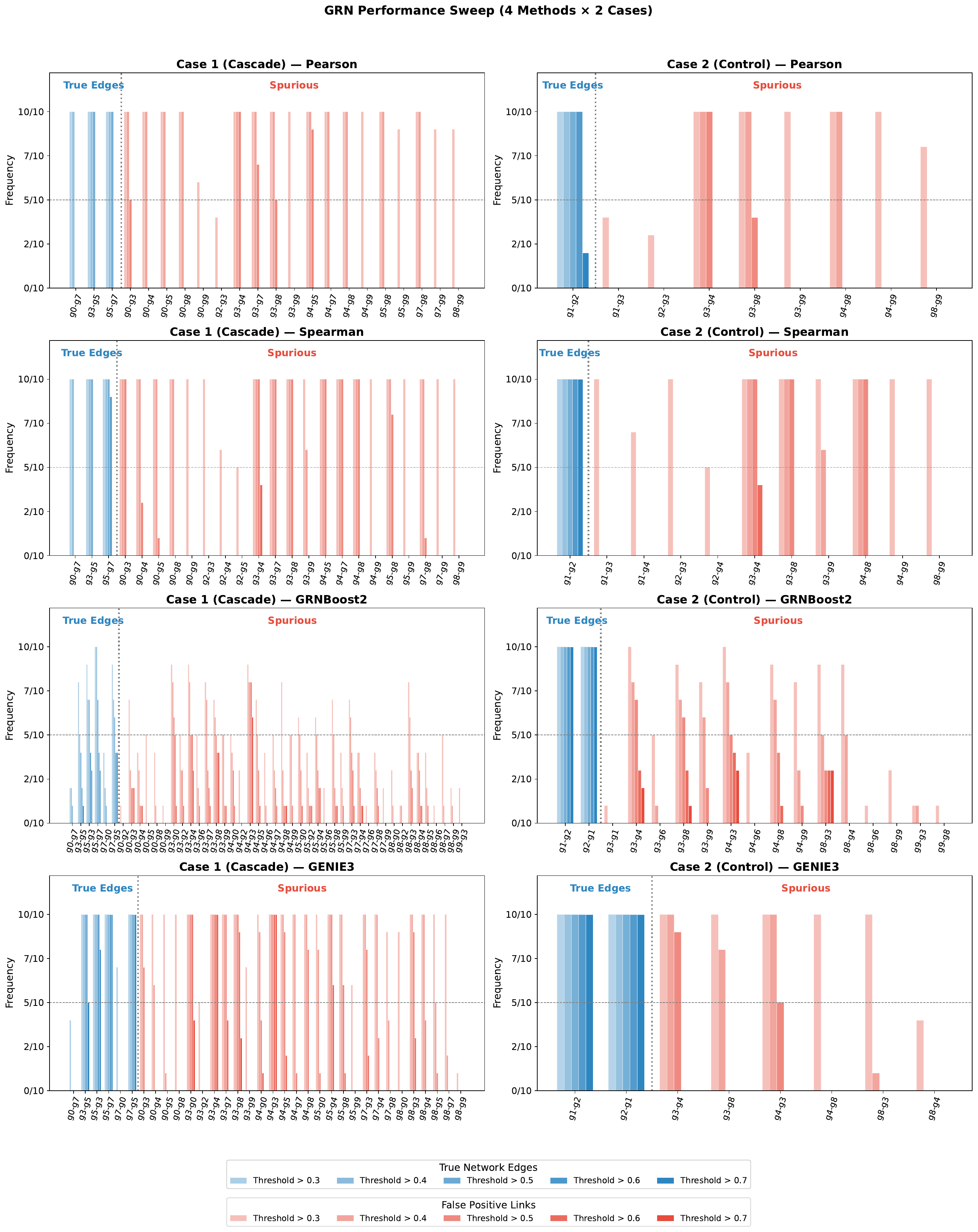}
  \caption{\textbf{GRN inference benchmark across four methods and two cases.}
    Edge frequency across 10 independently seeded replicates and five thresholds $\tau \in \{0.3,\,0.4,\,0.5,\,0.6,\,0.7\}$ (color intensity 
    encodes $\tau$; blue = programmed $CX$ edge, red = spurious edge). Rows correspond to inference methods (Pearson, Spearman, GRNBoost2, 
    GENIE3); columns correspond to cases. \textbf{Left column} (inter-state cascade, $L_1$): programmed edges 
    ($g_0$--$g_7$, $g_3$--$g_5$, $g_5$--$g_7$) are recovered at $\tau \leq 0.4$ but degrade at stricter thresholds; spurious edges 
    driven by $\theta_3{=}0.9\pi$ and $\theta_4{=}0.8\pi$ persist throughout for all methods. 
    \textbf{Right column} (non-interacting control, $L_2$): the $g_1$--$g_2$ link is recovered at $10/10$ replicates across all thresholds by 
    GRNBoost2 and GENIE3, and up to $\tau = 0.6$ by Pearson and Spearman; GENIE3 produces the fewest spurious edges (six total).}
  \label{fig:grn_corr}
\end{figure}

The GRN benchmark evaluates four inference methods (Pearson, Spearman, GRNBoost2, and GENIE3) on their ability to recover the programmed quantum entanglement topology from the synthetic data.
Following the preprocessing pipeline described in Section~\ref{sec:network}, network inference was restricted to the 10 genes modeled within the quantum 
kernel ($g_0$--$g_9$); housekeeping genes were excluded as their expression is not governed by the entanglement topology. 
For correlation-based methods, Pearson and Spearman matrices were computed on the processed expression matrix and thresholded at $|\text{Corr}| > \tau$; for tree-based methods, GRNBoost2 and GENIE3 importance scores were normalized to $[0,1]$ per replicate before applying the same threshold sweep, where $\tau \in \{0.3,\,0.4,\,0.5,\,0.6,\,0.7\}$. 
To assess stability, the full pipeline was repeated across 10 independently seeded replicates and edge frequency, the fraction of replicates in which an 
edge appears, was computed for every (method, case, threshold) combination. Results are summarized in Fig.~\ref{fig:grn_corr}.
\paragraph{Case~1 (inter-state cascade, $L_1 = \{q_3{\to}q_5{\to}q_7{\to}q_0\}$).}
Both Pearson and Spearman recover the programmed $CX$ edges ($g_0$--$g_7$, $g_3$--$g_5$, $g_5$--$g_7$) at low thresholds ($\tau \leq 0.4$), but these true edges degrade and disappear as $\tau$ increases. Spurious edges degrade as well with increasing threshold, yet remain present in higher proportion than the true edges, indicating a high false discovery rate in both methods. Spearman retains the $g_5$--$g_7$ link up to $\tau = 0.6$, showing slightly higher sensitivity than Pearson at stricter thresholds. The \emph{high-probability bias} driven by $\theta_3=0.9\pi$ and $\theta_4=0.8\pi$ systematically inflates spurious co-occurrences, making precise recovery of the true topology unfeasible \cite{wang2015efficient}.
\paragraph{Case~2 (non-interacting control, $L_2 = \{q_2{\to}q_1\}$).}
Both Pearson and Spearman recover the programmed $g_1$--$g_2$ link up to $\tau = 0.6$. The spurious $g_3$--$g_4$ edge persists across thresholds, again driven by the high base activation of those genes. Overall, Case~2 yields a cleaner network than Case~1, as the majority of genes have low activation angles and contribute fewer coincidental correlations, yet false discoveries remain present throughout.
\paragraph{Tree-based inference: GRNBoost2 and GENIE3 (Cases 1 and 2).}
Extending the benchmark to tree-based GRN methods reveals a qualitatively different recovery profile.
GRNBoost2 and GENIE3 produce directed importance scores, max-normalized to $[0,1]$ per replicate, before applying the same threshold sweep $\tau \in \{0.3,\,0.4,\,0.5,\,0.6,\,0.7\}$.
In Case~2 (non-interacting control, $L_2 = \{q_2{\to}q_1\}$), both methods recover the programmed $g_1$--$g_2$ interaction in both the forward ($g_2{\to}g_1$) and reverse ($g_1{\to}g_2$) directions at frequency $10/10$ across all replicates and all five thresholds, with GENIE3 producing markedly fewer spurious edges (six total) than GRNBoost2. 
This improved recovery arises because tree-based methods compute conditional variable importances rather than pairwise marginal statistics, making them a more natural match for the conditional dependency structure encoded by $CX$ gates.
In Case~1 (inter-state cascade, $L_1 = \{q_3{\to}q_5{\to}q_7{\to}q_0\}$), GRNBoost2 and GENIE3 recover the programmed cascade edges, including both forward and reverse directed forms of $g_0$--$g_7$, $g_3$--$g_5$, and $g_5$--$g_7$, at lower thresholds, confirming sensitivity to the multi-hop entangled structure.
The sensitivity advantage carries a precision cost, as the deliberately high activation angles $\theta_3=0.9\pi$ and $\theta_4=0.8\pi$ (Table~\ref{tab:initial_angles}) introduce structural co-expression confounders by design, and tree-based methods, being more sensitive, amplify both the true cascade signal and these designed spurious patterns simultaneously. 
True-edge recovery is lower than under Pearson or Spearman, with GRNBoost2 reaching $2$--$6/10$ and GENIE3 up to $7$--$10/10$ at $\tau \leq 0.4$. The high baseline activation of $g_3$ ($\theta_3=0.9\pi$) and $g_4$ ($\theta_4=0.8\pi$) generates coincidental co-expression patterns that any sensitive method will detect alongside the true cascade edges.
\paragraph{Interpretation.}
The consistent finding across all thresholds and replicates is that pairwise correlation methods recover true $CX$ edges only at lenient thresholds, where spurious edges also proliferate, and lose them at stricter thresholds where specificity improves. This trade-off confirms that both Pearson and Spearman are systematically biased toward high-expression genes, capturing emergent statistical co-dependencies rather than the underlying causal structure \cite{gonzalez2017random}. The $CX$ gates entangle qubits into a shared quantum state, so the resulting expression patterns reflect higher-order dependencies across the cascade that pairwise correlation, operating on gene pairs in isolation, cannot reconstruct.
Tree-based methods show substantially better alignment, particularly in Case~2, where full $10/10$ recovery with few false positives confirms that importance-based inference is better suited to the joint probability structure imposed by $CX$ gates than marginal correlation.
The Case~1 result is therefore not a failure of tree-based inference; both methods successfully detect the programmed cascade links, with GENIE3 approaching the recovery consistency of Pearson and Spearman at lenient thresholds while additionally resolving edge directionality. The elevated false-positive rate is a direct consequence of the deliberately high activation angles in Table~\ref{tab:initial_angles}; any sufficiently sensitive method will capture the designed confounders alongside the true structure. This precision-sensitivity trade-off is itself a reproducible, quantifiable property of the simulation ground truth, demonstrating that qSimCells can independently tune entanglement topology and activation-probability bias to stress-test GRN inference algorithms in a controlled, interpretable manner.

\subsubsection{Detecting cell-cell interactions with synthetic data}
We applied CellChat, a widely used classical method of cell-cell interaction detection, to infer cell-cell communication patterns from synthetic scRNA-seq data. CellChat predicts intercellular signaling by modeling interaction probabilities of LR pairs \cite{cellchat}. Our objective was to evaluate whether CellChat could accurately recover the true mechanistic interactions embedded within the simulated data while distinguishing them from spurious correlations driven by highly expressed but non-interacting genes.

In our experimental setup, the custom LR database contained two categories of interactions (1) True inter-state interactions ($g_3 \to g_5$ and $g_7 \to g_0$), which represent the mechanistic LR pairs responsible for genuine intercellular signaling; and (2) False control pairs ($g_8 \to g_4$ and $g_9 \to g_4$), which were included to evaluate CellChat's ability to discriminate between authentic and coincidental association. Details of these four simulated LR interactions are provided in Table \ref{tab:sim_cc_inter_db}.

\begin{table}[h!]
 \centering
 \caption{Simulated Ligand-Receptor interactions}
 \label{tab:sim_cc_inter_db}
 \begin{tabular}{cccccc}
 \toprule
 \textbf{Interaction Name} & \textbf{Pathway} & \textbf{Ligand} & \textbf{Receptor} & \textbf{Annotation} & \textbf{Evidence} \\
 \midrule
 g3\_g5\_simulated & Simulated1 & g3 & g5 & Secreted Signaling & Simulated1 \\
 g7\_g0\_simulated & Simulated2 & g7 & g0 & Secreted Signaling & Simulated2 \\
 g8\_g4\_simulated & Simulated3 & g8 & g4 & Secreted Signaling & Simulated3 \\
 g9\_g4\_simulated & Simulated4 & g9 & g4 & Secreted Signaling & Simulated4 \\
 \bottomrule
 \end{tabular}
\end{table}

The results presented in Table \ref{tab:cellchat_inference}, validate that the proposed gene activation patterns (from Table \ref{tab:initial_angles}) and the corresponding entanglement topologies ($L_1$ and $L_2$) provide strong mechanistic consistency with the designed regulatory framework. As anticipated, the true LR pairs exhibited substantial increase in inferred communication probabilities when the inter-state interactions were activated. For instance, the LR pair $g_3 \to g_5$ increased from 0.015 (Mo) to 0.141 (Co), representing a 9.5-fold enhancement (Ratio). Similarly, $g_7 \to g_0$ exhibited an even greater amplification from 0.001 (Mo) to 0.099 (Co), a 98-fold increase (Ratio) a pronounced activation of cross-cell signaling. Furthermore, the ligand $log_2FC = 1.607$ ($p_{DE} < 0.001$) and receptor $log_2FC = 1.558$ ($p_{DE} < 0.001$) confirm upregulation in Co vs. Mo.
These findings are fully consistent with our design principle, in which entanglement topologies were orchestrated by the master gene regulator, $g_3$, thereby validating the role of quantum entanglement in modeling mechanistic intercellular communication. 
These ratios are fully reproducible, as all stochastic components of the pipeline are globally seeded (see Section~\ref{sec_methods} and Supplemental S1), making the reported values deterministic. The shot-count convergence analysis (Supplemental Fig.~S1, Spearman $\rho \geq 0.93$) further confirms stability of the underlying quantum distribution across shot counts.

Conversely, the false LR pairs, which were included in the custom database to represent coincidentally co-expressed but non-mechanistic genes, showed minimal changes in inferred communication probability. This outcome aligns with expectations, as no direct entanglement (mechanistic link) was set for these pairs. For instance, the communication probability for $g_9 \to g_4$ showed only a marginal change, shifting from 0.077 (Mo) to 0.076 (Co), corresponding to a ratio of 0.98 (approximately unchanged). This result confirms that the inference method  effectively distinguished true mechanistic links from spurious co-activation driven by expression magnitude. Crucially, while the method still assigned high-confidence probabilities (all $p_{cc}<0.001$) even in the absence of mechanistic coupling, such absolute measures can be misleading. The relative change in communication probability between the interacting and no-interacting states, not the absolute probability or p-value, emerged as the more reliable indicator for reliably validating the synthetic mechanistic links \cite{storey2002direct}. This interpretation is further reinforced by the register architecture: since CT1 cells carry zero expression of $g_5$--$g_9$ and CT2 cells carry zero expression of $g_0$--$g_4$ by design, the probability shift observed in the Co condition cannot originate from coincidental co-expression, it is a direct readout of the inter-register entanglement topology.

\begin{table}[h!]
\centering
\caption{\textbf{CellChat ligand-receptor inference and cell-type-specific cross-condition differential expression.} 
Condition-specific communication metrics are presented as \emph{Probability} ($p_{\text{cc}}$, cell communication $p$-value), where \emph{Ratio} = Prob.(Co)/Prob.(Mo). 
To evaluate corresponding expression changes, independent gene-level differential expression (Co vs.\ Mo) was performed via a Wilcoxon rank-sum test for the ligand (L.) and receptor (R.) pairs and reported as $\log_2$FC ($p_{\text{DE}}$, differential expression $p$-value). 
Ligand statistics are calculated strictly within the sending (\emph{Source}) cell type, and receptor statistics are calculated strictly within the receiving (\emph{Target}) cell type.}
\label{tab:cellchat_inference}
\footnotesize
\setlength{\tabcolsep}{2 pt}
\begin{tabular}{llcccclll}
\toprule
\textbf{Sou.} & \textbf{Tar.} & \textbf{L.} & \textbf{R.} & \textbf{Prob. Mo ($p_{\text{cc}}$)} & \textbf{Prob. Co ($p_{\text{cc}}$)} & \textbf{Ratio} & \textbf{L.\ $\log_2$FC ($p_{\text{DE}}$)} & \textbf{R.\ $\log_2$FC ($p_{\text{DE}}$)} \\
\midrule
CT1 & CT2 & $g_3$ & $g_5$ & 0.015 ($<$0.001) & 0.141 ($<$0.001) & 9.50  & 0.014 (0.914)     & 1.711 ($<$0.001)   \\
CT2 & CT1 & $g_7$ & $g_0$ & 0.001 ($<$0.001) & 0.099 ($<$0.001) & 97.89 & 1.607 ($<$0.001)   & 1.558 ($<$0.001)   \\
CT2 & CT1 & $g_8$ & $g_4$ & 0.121 ($<$0.001) & 0.121 ($<$0.001) & 1.00  & 0.051 (0.423)     & $-$0.020 (0.474)   \\
CT2 & CT1 & $g_9$ & $g_4$ & 0.077 ($<$0.001) & 0.076 ($<$0.001) & 0.98  & 0.014 (0.899)     & $-$0.020 (0.474)   \\
\bottomrule
\end{tabular}
\end{table}

\section{Discussion}\label{sec_disc_n_conc}
To conclude, our work introduces qSimCells, a novel hybrid quantum-classical simulator developed to address the fundamental challenge of generating realistic scRNA-seq data. A recent benchmark study has demonstrated that classical methods struggle with complex designs and yield unreliable performance estimates \cite{shaky}, justifying the need for a new approach. 
To situate qSimCells within the broader landscape of single-cell simulators, we benchmarked it against two established classical tools, SERGIO~\cite{sergio} and scMultiSim~\cite{multisim}, across a GRN differentiation gradient of increasing regulatory strength ($t_1 \to t_4$, Supplemental Section~S2). All three simulators recover the programmed GRN cascade in GENIE3 AUROC, which rises monotonically with regulatory strength and reaches near-perfect recovery at $t_4$. A key architectural distinction emerges in the manifold geometry, however. In this benchmark circuit, qSimCells cells cluster into sharply discrete, well-separated islands in PHATE embeddings~\cite{moon2019visualizing} corresponding to the finite set of reachable binary gene-expression states, whereas SERGIO populations separate along a smooth expression gradient with continuous intra-cluster transitions driven by Langevin stochastic dynamics. This behaviour reflects Born-rule measurement collapse under the present single-angle $CRX(\theta)$ design; richer entanglement topologies and multi-parameter gate sequences would populate a larger fraction of the $2^N$ gene-state basis, producing denser Born-rule distributions and recovering progressively more continuous expression manifolds. More broadly, qSimCells occupies an adjacent rather than competing niche to SDE- or CIF-based simulators. It encodes gene regulatory interactions through quantum amplitudes and interference, producing joint probability distributions over gene-state combinations whose richness and continuity scale with circuit topology, a property natively inaccessible to classical probabilistic models of the same graph structure (Supplemental Section~S2). Quantum generative models are uniquely positioned to handle the complex and highly correlated probability distributions required for realistic biological data, leveraging the exponentially large Hilbert space of quantum registers \cite{kyriienko2024}.

By leveraging quantum entanglement, qSimCells encodes complex, nonlinear topologies of GRNs and cell-cell communication. To the best of our knowledge, qSimCells is the first quantum-based computational tool that simulates intracellular and intercellular dynamics in a natural and integrated manner. 
While CIF-based kinetic simulators such as scMultiSim~\cite{multisim} are well-suited for contact-dependent signaling between highly distinct cell types (such as immune-tumor interactions), qSimCells encodes cell-cell communication directly as quantum entanglement between gene-expression registers, a transcription-native and location-independent design that complements spatial approaches and is particularly promising for modeling transcriptional state transitions such as stem cell differentiation, where regulatory crosstalk between lineages is driven by gene-expression coupling rather than geometric proximity.
The central advantage of our quantum kernel lies in its ability to enforce correlated dependencies through the application of $CX$ gates \cite{nielsen}, thereby establishing an entangled hub, a joint quantum state in which the coupled genes form a physically inseparable system. The circuit topology encodes a directed graph as the simulator's known ground truth by construction; we note, however, that quantum entanglement itself is non-directional, and different $CX$ configurations can yield similar output distributions, so the programmed topology is a design choice rather than a quantity uniquely recoverable from samples alone. This enables the modeling of biologically meaningful entangled pathway hubs such as $\{g_3, g_5, g_7, g_0\}$, programmed in the circuit as $g_3 \to g_5 \to g_7 \to g_0$, whose joint correlation structure is not naturally expressible in classical correlation-based simulators \cite{shaky}. Concerning scalability, the human genome encodes over 20,000 protein-coding genes with more than one million autonomous exons~\cite{stepankiw2023human}, far exceeding current quantum hardware capacity, which supports tens to a few hundred qubits. However, qSimCells uses a modular design in which the quantum kernel captures the causal GRN core while the classical NB augmentation handles the full transcriptome (Section~\ref{subsec_count_mat}). A hybrid strategy, modeling a pathway of interest (10--50 genes) embedded within a classically simulated background, is the most practical approach for full-genome scale at current hardware.

Our results demonstrate that the synthetic datasets generated by qSimCells exhibit entanglement-enforced statistical dependencies, making them an ideal ground truth for benchmarking advanced inference algorithms. 
When correlation-based methods (Pearson, Spearman) were applied to the quantum-generated data, they failed to recover the programmed $CX$ topology at stricter thresholds, producing spurious associations driven by high-expression genes. Tree-based methods (GRNBoost2, GENIE3) showed substantially better alignment, with GENIE3 achieving full recovery in Case~2 (Fig.~\ref{fig:grn_corr}).
Correlation-based methods were particularly sensitive to baseline expression probabilities determined by initial gene activation angles ($\theta_i$), often overemphasizing correlations such as $g_3-g_4$ that overshadowed the true, entanglement-induced dependencies \cite{shaky}. These findings confirm that classical correlation-based approaches capture only pairwise marginals, prioritizing expression magnitude over the joint entanglement structure and therefore cannot reliably recover the true causal structure, a limitation that importance-based methods partially overcome.

Application of CellChat to the inter-state communication data provided strong validation of our quantum-derived ground truth while also revealing methodological limitations of cell-cell communication inference frameworks \cite{cellchat}. The tool successfully identified the true LR pairs ($g_3 \to g_5$ and $g_7 \to g_0$) with a 9.5- to 98-fold increase in communication probability under interacting conditions, consistent with the programmed entanglement. However, the false LR pairs, which lacked mechanistic links, were still assigned high-confidence probabilities (e.g., $p_{cc} < 0.001$). Such absolute confidence, in the absence of mechanistic coupling, can be misleading unless compared across experimental conditions. Therefore, our results highlight that the relative change in communication probability-between interacting and non-interacting state, serves as a more reliable indicator of true causal signaling than absolute probability or p-values.

Critically, we validated that qSimCells circuits produce consistent expression distributions across simulated and physical quantum hardware. Executing the co-culture and mono-culture circuits on IBM Marrakesh (127-qubit Eagle r3 processor) under four conditions, ideal noiseless simulation, IBM noise-model simulation, real hardware raw, and real hardware with error mitigation (dynamical decoupling XY4~\cite{viola1999dynamical}, Pauli gate twirling~\cite{knill2008randomized}, and TREX~\cite{vandenberg2022modelfree,vandenberg2023probabilistic}), we find that the rank ordering of per-gene $P(\text{gene}=1)$ is preserved across all hardware conditions (Spearman $\rho > 0.98$ relative to the ideal simulation, Supplemental Section~S4, Fig.~S3). The noisy simulator closely predicts the raw hardware results, confirming that IBM's noise model is an accurate characterisation of the device at this circuit depth. Error mitigation further reduces RMSE relative to the raw hardware run, demonstrating that shallow qSimCells circuits are already robust to realistic gate and readout noise on current NISQ devices.

In conclusion, qSimCells represents a quantum-enhanced platform for generating high-fidelity synthetic scRNA-seq data. Through quantum entanglement, it establishes a controlled and interpretable ground truth characterized by entanglement-enforced correlations, nonlinearity and a programmable causal topology not naturally encoded in classical correlation-based generative models. This framework not only exposes the inherent weakness of traditional inference techniques but also offers a new paradigm for developing algorithms capable of capturing the complex structural dependencies intrinsic to gene regulation. 
Ultimately, our results confirm that the quantum kernel is essential for constructing benchmark datasets in which the generative topology is unequivocally defined by construction, paving the way for next-generation computational methods that move beyond linear correlations toward a truly mechanistic understanding of cellular systems.

A natural extension of qSimCells is its transformation into a fully learned generative model. The gene activation angles $\theta_i$ can be directly estimated from observed single-cell expression frequencies, providing an immediate, parameter-free calibration against real data. More generally, replacing the $CX$ gates with parametrized $CRX$ gates yields a trainable quantum circuit whose parameters can be optimized by minimizing a Kullback-Leibler divergence-based loss function, following the hybrid quantum-classical framework established in qscGRN~\cite{qgrn}. This would transform qSimCells into a data-driven generative model capable of learning complex gene regulatory topologies directly from single-cell profiles. However, this approach introduces well-known challenges: classical optimizers navigating quantum loss landscapes are susceptible to barren plateaus and local minima, which may prevent the learned circuit from faithfully representing the underlying data distribution. Addressing these challenges represents an important direction for future work.

We note that the present circuit family, $R_y$ rotations on $\ket{0}$ followed by $CX$ gates, keeps all amplitudes real and non-negative, and the resulting distribution is classically reproducible in $O((n+n')M)$ time. The advantage of qSimCells is therefore architectural rather than computational: the PQC provides a compact, principled encoding of programmed entanglement topologies not naturally available in classical correlation-based simulators.

Finally, our findings argue for a quantum-native, generative paradigm in machine learning \cite{kyriienko2024,gao2022enhancing}, particularly for single-cell modeling. Quantum circuits naturally encode joint probability distributions through superposition and entanglement, making them well-suited to synthesize transcriptomes that embed nonlinear, causal gene-gene and cell-cell dependencies. qSimCells exemplifies this advantage. In contrast, quantum machine learning that simply mirrors classical discriminative pipelines and losses—designed for classical hardware—fails to exploit quantum advantages and often incurs prohibitive readout and training overhead. Thus, generative quantum models should be the primary path forward, coupled with quantum-native objectives and inference strategies capable of capturing the complex causal dependencies inherent in gene regulation.

\backmatter

\section*{Supplementary information}
All required materials are included within the manuscript, supplemental information and GitHub.

\section*{Acknowledgements}

\section*{Declarations}
\begin{itemize}
\item\textbf{Funding}: This study was funded by the U.S. Department of Defense (DoD, GW200026) and the National Institute for Environmental Health Sciences (NIEHS, P30 ES029067) for JJC, Allen Endowed Chair in Nutrition \& Chronic Disease Prevention for RSC, and the Cancer Prevention \& Research Institute of Texas (CPRIT, RP230204) for JJC and RSC.\\
\item\textbf{Conflict of interest}: The authors have no conflict of interest.\\
\item\textbf{Data availability}: Our algorithm is publicly available on GitHub \url{https://github.com/cailab-tamu/qSimCells}.\\
\item\textbf{Author contribution}: SR contributed to conceptualization, implemented the software, developed the method, and led the writing. VSK prepared the data and contributed to writing. RSC provided funding and supervision and contributed to writing. JJC conceived the concept, co-developed the method, contributed to writing, and secured funding.
\end{itemize}

\bibliography{sn-bibliography}

\clearpage
\setcounter{figure}{0}
\setcounter{table}{0}
\setcounter{equation}{0}
\setcounter{section}{0}
\renewcommand{\thefigure}{S\arabic{figure}}
\renewcommand{\thetable}{S\arabic{table}}
\renewcommand{\theequation}{S\arabic{equation}}
\renewcommand{\thesection}{S\arabic{section}}
\renewcommand{\thesubsection}{S\arabic{section}.\arabic{subsection}}

\setcounter{figure}{0}
\setcounter{table}{0}
\setcounter{equation}{0}
\setcounter{section}{0}
\renewcommand{\thefigure}{S\arabic{figure}}
\renewcommand{\thetable}{S\arabic{table}}
\renewcommand{\theequation}{S\arabic{equation}}
\renewcommand{\thesection}{S\arabic{section}}
\renewcommand{\thesubsection}{S\arabic{section}.\arabic{subsection}}

\begin{center}
{\LARGE\textbf{Supplementary Information}}\\[0.5em]
{\large Quantum Generative Modeling of Single-Cell Transcriptomes:\\
Capturing Gene--Gene and Cell--Cell Interactions}
\end{center}
\vspace{1em}

\section{Reproducibility and Seeding}
\label{sec:S1}

All stochastic operations in the qSimCells pipeline are globally seeded to ensure full numerical reproducibility.  
The Qiskit \texttt{AerSimulator} is instantiated with \texttt{seed\_simulator=42} via the \texttt{SamplerV2} primitive options; NumPy random state is set with \texttt{np.random.seed(42)} at the start of every replicate loop; Scanpy UMAP (Uniform Manifold Approximation and Projection) and Leiden clustering use \texttt{random\_state=42} at every call; and the CellChat permutation null is seeded via R's \texttt{set.seed(123)}.  
The classical benchmark simulators follow the same convention: the scMultiSim R script uses \texttt{rand.seed=42} (co-culture) and \texttt{rand.seed=43} (mono-culture), while SERGIO uses \texttt{np.random.seed(42)}.  
All code, together with a pinned \texttt{environment.yml} and per-figure run instructions, is publicly available at \url{https://github.com/cailab-tamu/qSimCells}.

\section{Shot-Count Convergence Analysis}
\label{sec:S2}

\subsection{Motivation}

The main-text Fig.~2 histograms were generated with $N_{\text{shots}} = 2{,}000$ over a 5-qubit register that has $2^5 = 32$ possible basis states, yielding an average of only $\approx 62$ counts per outcome.  
Values such as 248, 130, and 162 are therefore noisy point estimates.  
This section verifies that the qualitative conclusions drawn from the 2{,}000-shot histogram are preserved at $N_{\text{shots}} = 100{,}000$ and provides Wilson score confidence intervals for every reported probability.

\subsection{Method}

The two entanglement configurations from the main text were re-simulated at $N_{\text{shots}} = 100{,}000$ (seed = 42) using the same \texttt{AerSimulator} backend, the same rotation angles (Table~1 of the main text), and the same $CX$ topologies: $L_1 = \{(3,5),(5,7),(7,0)\}$ (co-culture, inter-state cascade) and $L_2 = \{(2,1)\}$ (mono-culture, non-interacting control).

Let $N = 100{,}000$ be the total shot count and $c_s$ the raw count for basis state $s \in \{0,\ldots,31\}$.  
The estimated outcome probability is $\hat{p}_s = c_s / N$.  
The 95\% Wilson score confidence interval is:

\begin{equation}\label{eq:wilson}
    \hat{p}_s \;\pm\; \frac{1.96\,\sqrt{\hat{p}_s(1-\hat{p}_s)/N +
    1.96^2/(4N^2)}}{1 + 1.96^2/N}.
\end{equation}

For the dominant outcomes (e.g.\ $\hat{p}_s \approx 0.15$) the half-width is $\approx 0.002$, confirming that $N = 10^5$ shots yield well-converged estimates.

\subsection{Results}

\begin{figure}[h!]
    \centering
    \includegraphics[width=0.95\textwidth]{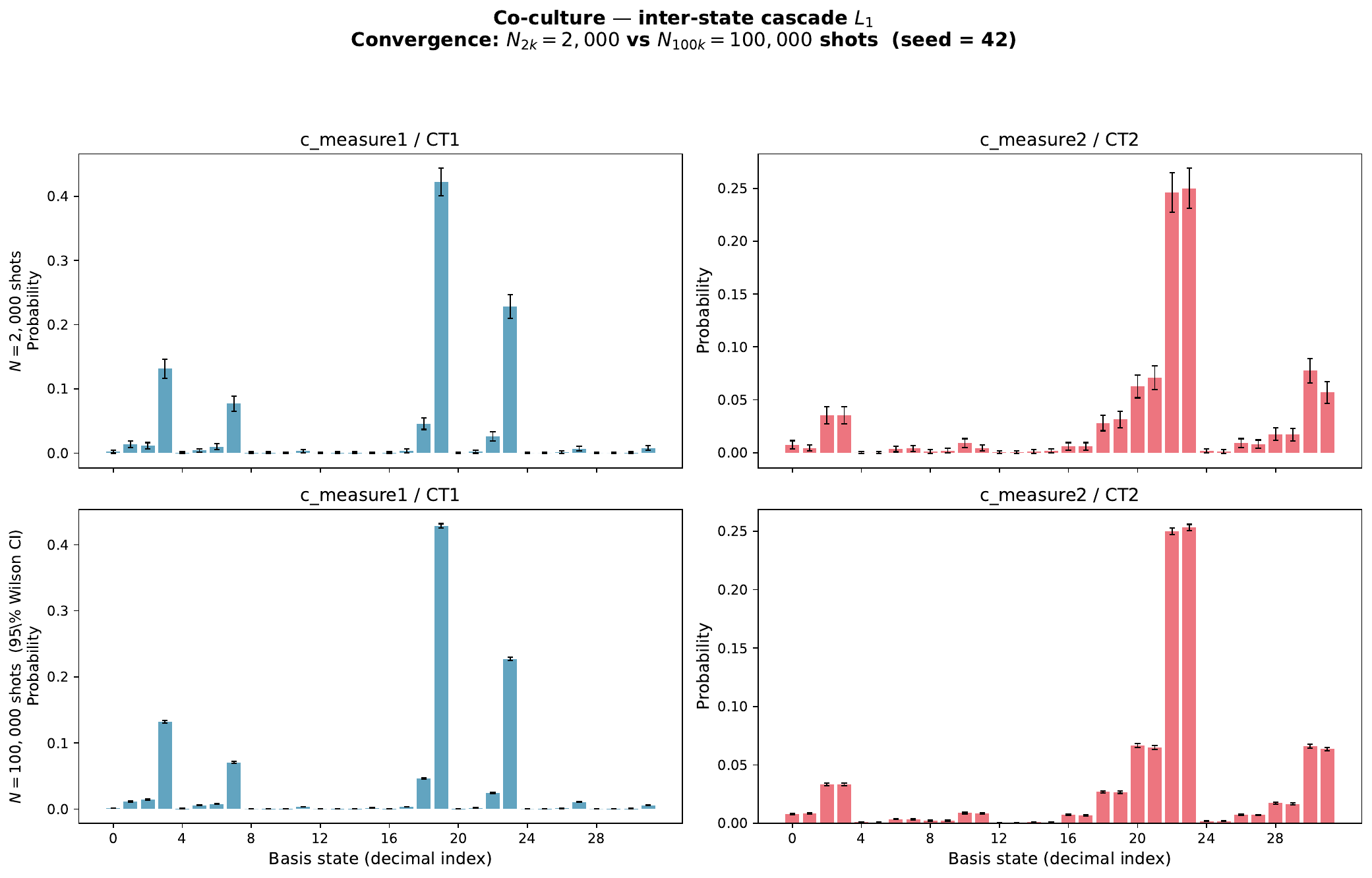}
    \caption{\textbf{Shot-count convergence: co-culture $L_1$, both registers.}
    Top row: $N=2{,}000$ shots; bottom row: $N=100{,}000$ shots.
    Left column: \texttt{c\_measure1} (CT1, cell type 1, blue); right column:
    \texttt{c\_measure2} (CT2, cell type 2, red).  Error bars show 95\% Wilson CI in both rows;
    at $N=100{,}000$ they are nearly invisible, confirming convergence.
    The distribution shape is visually identical across rows, and the rank ordering
    is preserved (Spearman $\rho = 0.9324$, $p = 8.47\times10^{-15}$),
    consistent with the probability distribution being natively encoded in the
    quantum kernel and faithfully recovered by sampling.}
    \label{fig:convergence_combined}
\end{figure}

Fig.~\ref{fig:convergence_combined} shows the measurement histograms for the co-culture topology $L_1$ at both $N=2{,}000$ and $N=100{,}000$ shots with 95\% Wilson confidence intervals on both rows.  
At $N=100{,}000$ the error bars are smaller than the bar width for all dominant outcomes, while at $N=2{,}000$ they are visibly larger, the contrast makes the convergence argument graphically immediate.

A key design principle of qSimCells is that the probability distribution over basis states is \emph{natively embedded within the quantum kernel}: the $R_y$ rotation angles encode gene-activation probabilities directly into the quantum amplitudes, and the $CX$ entanglement gates propagate conditional dependencies across registers.  
Consequently, the Born-rule measurement distribution is not a post-hoc approximation, it \emph{is} the generative model.  
Both theory (the parameterized circuit defines an exact probability distribution over $2^n$ outcomes) and practice (Fig.~\ref{fig:convergence_combined}, Spearman $\rho = 0.9324$ at $N = 2{,}000$ vs $N = 100{,}000$) confirm that sampling from this kernel faithfully recovers the programmed distribution, with confidence intervals shrinking as $\mathcal{O}(1/\sqrt{N})$ as expected for independent Bernoulli trials.

\subsection{Convergence table}

Table~\ref{tab:shot_convergence} reports raw counts and outcome probabilities for the top-5 basis states of the CT1 register at both $N = 2{,}000$ and $N = 100{,}000$ shots.  
Including the absolute counts at both shot levels makes the sampling variability tangible: e.g.\ State~19 is observed 844 times out of 2{,}000 (42.20\%) and 42{,}878 times out of 100{,}000 (42.88\%), a difference well within the Wilson CI.  
The Spearman rank correlation $\rho = 0.9324$ ($p = 8.47\times10^{-15}$) over all 32 outcomes confirms that the 2k run already captures the converged rank ordering of the full distribution.

\begin{table}[h!]
    \centering
    \caption{Top-5 most probable basis states at $N=2{,}000$ and $N=100{,}000$ shots,
    co-culture inter-state cascade $L_1$, both registers (5 qubits each, 32 states).
    ``State'' = decimal index of the 5-bit measurement string;
    e.g.\ CT1 State~$19 = 10011_2$ corresponds to qubits $\{q_0, q_1, q_4\}$ in $|1\rangle$,
    driven by the high-angle qubit $q_4$ ($\theta_4 = 0.8\pi$) and the cross-register
    cascade $q_3{\to}q_5{\to}q_7{\to}q_0$.
    Count columns: raw counts out of $N=2{,}000$ and $N=100{,}000$ shots respectively.
    95\% CI: Wilson half-width at $N=100{,}000$ (Eq.~\ref{eq:wilson}).
    Spearman $\rho$ computed over all 32 outcomes per register.}
    \label{tab:shot_convergence}
    \begin{tabular}{cccccc}
        \toprule
        & \multicolumn{2}{c}{\textbf{Count}} & \multicolumn{2}{c}{\textbf{Freq.\ (\%)}} & \\
        \cmidrule(lr){2-3}\cmidrule(lr){4-5}
        \textbf{State} & \textbf{2k} & \textbf{100k} & \textbf{2k} & \textbf{100k} & \textbf{95\% CI ($\pm$\%)} \\
        \midrule
        \multicolumn{6}{c}{\textit{CT1 (\texttt{c\_measure1})}} \\
        \midrule
        19 & 844 & 42{,}878 & 42.20 & 42.88 & 0.307 \\
        23 & 456 & 22{,}719 & 22.80 & 22.72 & 0.260 \\
         3 & 263 & 13{,}204 & 13.15 & 13.20 & 0.210 \\
         7 & 154 &  7{,}062 &  7.70 &  7.06 & 0.159 \\
        18 &  92 &  4{,}630 &  4.60 &  4.63 & 0.130 \\
        \multicolumn{6}{l}{\small Spearman $\rho = 0.9324$ \quad ($p = 8.47\times10^{-15}$)} \\
        \midrule
        \multicolumn{6}{c}{\textit{CT2 (\texttt{c\_measure2})}} \\
        \midrule
        23 & 500 & 25{,}324 & 25.00 & 25.32 & 0.270 \\
        22 & 492 & 25{,}012 & 24.60 & 25.01 & 0.268 \\
        20 & 125 &  6{,}659 &  6.25 &  6.66 & 0.155 \\
        30 & 155 &  6{,}602 &  7.75 &  6.60 & 0.154 \\
        21 & 142 &  6{,}485 &  7.10 &  6.48 & 0.153 \\
        \multicolumn{6}{l}{\small Spearman $\rho = 0.9676$ \quad ($p = 1.72\times10^{-19}$)} \\
        \bottomrule
    \end{tabular}
\end{table}

\section{Classical Simulator Benchmark}
\label{sec:S3}

\subsection{Simulator architecture}
\label{sec:S3-arch}

Table~\ref{tab:sim_compare} summarizes the key architectural differences between the three simulators evaluated in this study.

\begin{table}[h!]
\centering
\caption{Architectural comparison of SERGIO, scMultiSim, and qSimCells.
``Intracellular GRN'' refers to a directed gene regulatory network within a single
cell type; ``Intercellular'' refers to a mechanism coupling expression \emph{across}
cell types.}
\label{tab:sim_compare}
\begin{tabular}{lccc}
\toprule
\textbf{Feature} & \textbf{SERGIO} & \textbf{scMultiSim} & \textbf{qSimCells} \\
\midrule
Intracellular GRN       & SDE (Langevin)        & CIF + Beta-Poisson     & PQC ($R_y$+$CX$)          \\
Intercellular           & None                  & Optional spatial CCI   & Entangled cross-register  \\
LR mechanism            & None                  & Proximity / signalling & Direct qubit entanglement \\
Cross-type cascade      & No                    & No                     & Yes ($L_1$)               \\
Stochastic count model  & Poisson               & Beta-Poisson           & Negative Binomial         \\
\bottomrule
\end{tabular}
\end{table}

SERGIO simulates intracellular GRN dynamics via a Langevin stochastic differential equation (SDE) in which transcription factor concentrations drive target gene production rates through a Hill-function activation term.  
SERGIO natively supports multiple cell types through per-type transcription rate profiles (\texttt{MrProfile}), and all cell types share the same GRN topology; however, GRN edges are strictly intra-cellular: a G4$\to$G5 edge means G4 regulates G5 within the same cell, not across cells.  
There is no inter-cellular signaling mechanism: each cell-type population is simulated independently, with no cross-type regulatory coupling.  
SERGIO therefore produces realistic intra-type co-expression patterns but cannot encode cell-to-cell GRN signals at the data-generation level.

scMultiSim is a CIF-based (Cell Identity Factor) kinetic simulator.  
Continuous latent variables propagate along a lineage tree and parameterize the burst frequency and size of a Beta-Poisson kinetic model for each gene.  
A GRN can be supplied as a static influence on the CIF-to-expression mapping.  
An optional spatial cell-cell interaction (CCI) module allows ligand--receptor (LR) pairs to modulate a neighboring cell's CIF values based on geometric proximity; this module was excluded from the present benchmark (see Section~\ref{sec:S3-impl}). 
Critically, neither the GRN module nor the spatial CCI module encodes a direct gene-level regulatory cascade spanning cell types.

\subsection{Motivation and scope}

This benchmark serves two purposes.  
First, it validates the GRN recovery result: if GRN inference also fails on data from purely classical simulators with the same ground-truth topology, then the failure is attributable to the intrinsic difficulty of recovering directed causal edges from correlation- or importance-based methods; not to the quantum data-generation mechanism.  
Second, it isolates the properties unique to qSimCells: the only simulator in the comparison that encodes a cross-cell-type directed gene regulatory cascade via quantum entanglement.

\subsection{Simulators}

All three simulators reproduce the same computational design: a single cell type (TypeA) is simulated at four GRN regulatory strengths (pseudo-timepoints $t_1 \to t_4$), yielding 500 cells per level and 2{,}000 cells in total per simulator.
Each simulator requires an internal workaround to satisfy its own API constraints while producing this single-population output, as described below.

\paragraph{qSimCells (this work).}
Parameterized quantum circuit (PQC) with $R_y$ activation gates and $CX$ ($CRX(\pi)$, a controlled rotation by $\pi$ around the $X$-axis) entanglement gates, encoding the intracellular GRN within a single unified kernel.  
The circuit contains two 5-qubit registers (CT1, CT2) with identical angles and GRN topology; no inter-register $CRX$ gates are present in this benchmark, as cross-type entanglement via topology $L_1$ is the subject of the main-text co-culture analysis.  
Since the two registers are statistically independent here, 500 shots of the circuit yield 500 CT1 measurements (\texttt{c\_measure1}) and 500 CT2 measurements; only the CT1 register is retained, giving 500 cells per regulatory level, consistent with SERGIO and scMultiSim.  
The $CRX(\theta)$ gate applies a controlled $R_X(\theta)$ rotation: at $\theta = \pi$ this yields the $CX$ gate used throughout; sweeping $\theta \in (0, \pi)$ introduces complex amplitudes into the quantum state, enabling interference effects that shape the probability distribution over all $2^5 = 32$ measurement outcomes in ways that have no analogue in classical SDE or CIF simulators.

\paragraph{SERGIO v2.}
SDE-based simulator with Langevin dynamics and Hill-function activation.  
SERGIO natively supports multiple cell types by assigning per-type transcription rate profiles (\texttt{MrProfile}); all types share the same GRN topology.  
Importantly, GRN edges are strictly \emph{intra-cellular}: a G4$\to$G5 edge means G4 regulates G5 within the same cell via the Hill function, not across cells.  
There is no inter-cellular signaling mechanism; TypeA and TypeB populations are simulated independently with no cross-type regulatory coupling.
Both TypeA (higher $G_0$ production rate) and TypeB (lower $G_0$ production rate) were generated (500 cells each); only TypeA cells were retained for the GRN gradient benchmark to ensure a fair single-type comparison across all three simulators.  
Implemented via the \texttt{sergio\_rs} Python package (Rust reimplementation, $\sim$150$\times$ faster).

\paragraph{scMultiSim.}
CIF (Cell Identity Factor) model with Beta-Poisson burst kinetics propagated through a bifurcating lineage tree.  
scMultiSim requires at least two leaf nodes in the lineage tree; here, 500 cells were generated as two pseudo-populations of 250 each (\texttt{discrete.pop.size=c(250,250)}) with \texttt{diff.cif.fraction}=0.10 to minimize lineage divergence, making the two populations effectively homogeneous.
All 500 cells were retained per regulatory level, consistent with the 500-cell count used for SERGIO and qSimCells.
The \texttt{diff.cif.fraction}=0.10 setting introduces minor between-population CIF variation as a technical requirement of the tree structure; this may produce a residual cluster split in UMAP/PHATE embeddings at low GRN strengths (t1), but the primary measured quantity, the AUROC trajectory rising with regulatory strength (t1$\to$t4), remains valid and comparable across all three simulators.
scMultiSim supports optional spatial cell-cell interaction (CCI) through a ligand-receptor (LR) module that modulates CIF values based on cellular proximity; this module acts as an additive layer on top of the GRN simulation and was not used here.  
Implemented via the official R package (\texttt{ZhangLabGT/scMultiSim}).

\subsection{Shared ground-truth GRN}

All three simulators used the same five-gene directed acyclic graph:

\begin{equation}\label{eq:grn_benchmark}
    G_0 \;\xrightarrow{\;+\;}\; G_1 \;\xrightarrow{\;+\;}\; G_2
    \;\xrightarrow{\;+\;}\; G_3, \qquad G_4 \text{ (independent control).}
\end{equation}

$G_0$ is the master regulator; $G_4$ has no regulatory connections and serves as a negative-edge control for AUROC (area under the receiver-operating-characteristic curve) evaluation.
All edges are activating.
Fifty housekeeping genes (HKGs, genes with stable expression across regulatory levels) were appended to every count matrix ($\mu_{\text{HKG}}=80$, $r_{\text{HKG}}=6$, identical across all three simulators); HKGs are excluded from GRN inference but retained for UMAP and NB-fit evaluation.

\subsection{Evaluation metrics}

GRN inference performance is quantified by the area under the receiver-operating-characteristic curve (AUROC) for GENIE3 importance scores against the ground-truth DAG (directed acyclic graph) (Eq.~\ref{eq:grn_benchmark}); AUROC = 0.5 is uninformative, 1.0 is perfect.
Negative-binomial (NB) marginal fits ($\hat{\mu}, \hat{r}$ via MLE (maximum likelihood estimation)) assess count-statistic plausibility; NB log-likelihood is computed per gene per level and averaged across $G_0$--$G_4$, then min-max normalized to 0--100\% within each simulator to allow visual comparison across simulators with different native count scales.

\subsection{Implementation notes}
\label{sec:S3-impl}

\paragraph{scMultiSim.}
scMultiSim v1.2 requires $\geq 2$ leaf nodes; the minimum safe value for \texttt{num.cifs}=20 is \texttt{diff.cif.fraction}=0.10 (exactly 2 differential CIFs), giving a two-leaf tree \texttt{((R1:1,R2:1):1)} whose leaves represent the same effective cell type. \texttt{discrete.cif=TRUE} with \mbox{\texttt{discrete.pop.size=c(250,250)}} assigns CIFs per-population; \texttt{scale.s=1.0} leaves kinetic scaling at its default. HKGs are generated in R with \texttt{set.seed(242)}. The spatial CCI module was excluded to ensure a fair comparison  with SERGIO, which does not model intercellular signalling.

\paragraph{Gene naming.}
All simulators use 0-indexed names ($G_0$--$G_4$) matching the qSimCells qubit convention; ground-truth edges $G_0{\to}G_1$, $G_1{\to}G_2$, $G_2{\to}G_3$ are all activating, and $G_4$ has no edges.

\subsection{Results}
\label{sec:S3-expected}

\subsubsection{GRN gradient computational design}
\label{sec:S3-gradient}

The benchmark uses a \emph{GRN differentiation gradient}: a single cell type (TypeA) is simulated at four regulatory strengths (pseudo-timepoints $t_1 \to t_4$), producing a trajectory in expression space driven solely by interaction strength.  
Table~\ref{tab:grn_params} gives the exact parameter values per simulator.

\begin{table}[h!]
\centering
\caption{GRN gradient parameter values across the four levels ($t_1$--$t_4$) for each
simulator.  All three simulators use the same ground-truth DAG
(Eq.~\ref{eq:grn_benchmark}).  Higher values correspond to stronger GRN cascade
activation in all cases.}
\label{tab:grn_params}
\begin{tabular}{lcccc}
\toprule
\textbf{Simulator / parameter} & $t_1$ & $t_2$ & $t_3$ & $t_4$ \\
\midrule
SERGIO; Hill $k$ (interaction amplitude)     & 0.01  & 0.20  & 2.0   & 20.0  \\
qSimCells; $CRX$ angle $\lambda$ ($\times\pi$) & 0.1   & 0.4   & 0.7   & 1.0   \\
scMultiSim; edge \texttt{effect} weight      & 0.001 & 0.20  & 1.00  & 3.00  \\
\bottomrule
\end{tabular}
\end{table}

Each level comprises $N_{\text{cells}} = 500$ cells; the four levels are concatenated to give 2\,000 cells per simulator.  
All simulations are seeded with \texttt{SEED}=42.  
The ground-truth GRN contains 3 directed activating edges ($G_0{\to}G_1$, $G_1{\to}G_2$, $G_2{\to}G_3$); all remaining 17 directed pairs (out of $5 \times 4 = 20$ off-diagonal entries) are zero, including all edges involving $G_4$; the GENIE3 AUROC baseline is 0.5.
Fifty housekeeping genes (HKG\textsubscript{0}--HKG\textsubscript{49}, NB $\mu_{\text{HKG}}=80$, $r_{\text{HKG}}=6$) are appended identically to all matrices to provide a common anchor for UMAP/PHATE (Potential of Heat-diffusion for Affinity-based Trajectory Embedding) \cite{moon2019visualizing} and NB-fit evaluation.

\subsubsection{Simulator parameters and design equivalences}
\label{sec:S3-params}

Table~\ref{tab:sim_equiv} maps the conceptually equivalent design choices across the three simulators, clarifying how each encodes the master regulator, cell-to-cell noise, and GRN strength.

\begin{table}[h!]
\centering
\caption{Design equivalences across SERGIO, qSimCells, and scMultiSim.
Each row names a conceptual design choice and shows how it is realized in each simulator.
``MrProfile'' = \texttt{sergio\_rs.MrProfile}; ``CIF'' = Cell Identity Factor.}
\label{tab:sim_equiv}
\begin{tabular}{p{3.5cm}p{3.5cm}p{3.5cm}p{3.5cm}}
\toprule
\textbf{Design concept}    & \textbf{SERGIO}           & \textbf{qSimCells}            & \textbf{scMultiSim}           \\
\midrule
GRN mechanism              & Hill ODE (ordinary differential equation, sigmoidal, biochemical) & $CRX$ quantum gate (entanglement amplitude) & CIF linear propagation (statistical weight) \\
\addlinespace
GRN strength parameter     & Hill $k$ (interaction effect amplitude) & $CRX$ angle $\lambda$ & \texttt{effect} (edge coupling weight) \\
\addlinespace
Master regulator $G_0$     & MrProfile basal $\in [1.20, 1.80]$; constitutively produced & $R_y(0.70\pi) \approx 79\%$ ON; near-constant & CIF-driven; mean expression rises with \texttt{effect} \\
\addlinespace
Cell-to-cell noise         & Langevin $\sigma=0.5$ (SDE noise term) & NB overdispersion $R_\text{vec}$ & \texttt{cif.sigma}=1.0 (CIF variability) \\
\addlinespace
Count output type          & Continuous SDE floats $\to$ NB sampled & Binary bitstring $\to$ NB augmented & Integer counts (kinetic burst model) \\
\addlinespace
$G_4$ (independent control)& Poisson$(1)$; low, below $G_0$ & $R_y(0.35\pi)$; low probability & CIF-driven 5th gene (no GRN edges) \\
\addlinespace
Equivalence workaround     & $n_{\text{types}}=2$; keep TypeA half & Same circuit twice; keep \texttt{c\_measure1} & \texttt{diff.cif.fraction}=0.10; effectively mono-culture \\
\bottomrule
\end{tabular}
\end{table}

\subsubsection{Pros and cons of each simulator}
\label{sec:S3-proscons}

\paragraph{SERGIO (SDE / Hill kinetics).}
Strengths: mechanistically grounded; explicit Hill-function cooperativity encodes sigmoidal activation curves; cascade directionality is faithfully preserved in the SDE dynamics.
Limitations: GRN edges are strictly intra-cellular: the Hill-function interactions describe gene-gene regulation within a single cell, with no mechanism for one cell's expression to influence another cell's production rates; TypeA and TypeB populations are simulated independently and only TypeA is retained, so cross-type GRN signals cannot be encoded at the data-generation level.  Count scale is in continuous SDE units requiring downstream NB augmentation.  At very high $k$ the SDE enters multi-stable regimes producing bimodal distributions per gene (discussed below).

\paragraph{qSimCells (PQC, Parameterized Quantum Circuit / quantum kernel).}
Strengths: intra- and intercellular GRN encoded in a single unified circuit; per-gene expression probabilities are encoded directly in the quantum amplitudes via the Born rule and used as success parameters for explicit NB count sampling; the $CRX$ (Controlled-RX gate) entanglement angle $\lambda$ provides a directly interpretable GRN strength knob; circuit topology maps cleanly to the ground-truth DAG.
Limitations: the binary-to-continuous count mapping (NB augmentation) introduces an additional modelling layer; constitutive $G_0$ at $R_y(0.70\pi)$ has low cell-to-cell variance, limiting GENIE3's ability to detect the $G_0{\to}G_1$ edge (discussed below); quantum simulation cost scales with shot count.

\paragraph{scMultiSim (CIF / Beta-Poisson kinetics).}
Strengths: integer counts from a biologically motivated burst kinetics model; directly outputs realistic sparse single-cell count matrices; CIF latent space separates lineage identity from GRN regulation.  
Limitations: the CIF baseline (\texttt{cif.mean}) must be set low (1.5 rather than the default 4.0) to prevent the CIF-driven expression floor from overwhelming the GRN signal; requires $\geq 2$ leaf nodes in the lineage tree, necessitating a two-population workaround even for single-population analyses; GRN edge coupling through the CIF space is statistical rather than mechanistic, so it does not enforce the same deterministic cascade backbone as SERGIO.

\subsubsection{Key computational observations}
\label{sec:S3-obs}

\begin{figure}[h!]
    \centering
    \includegraphics[width=\textwidth]{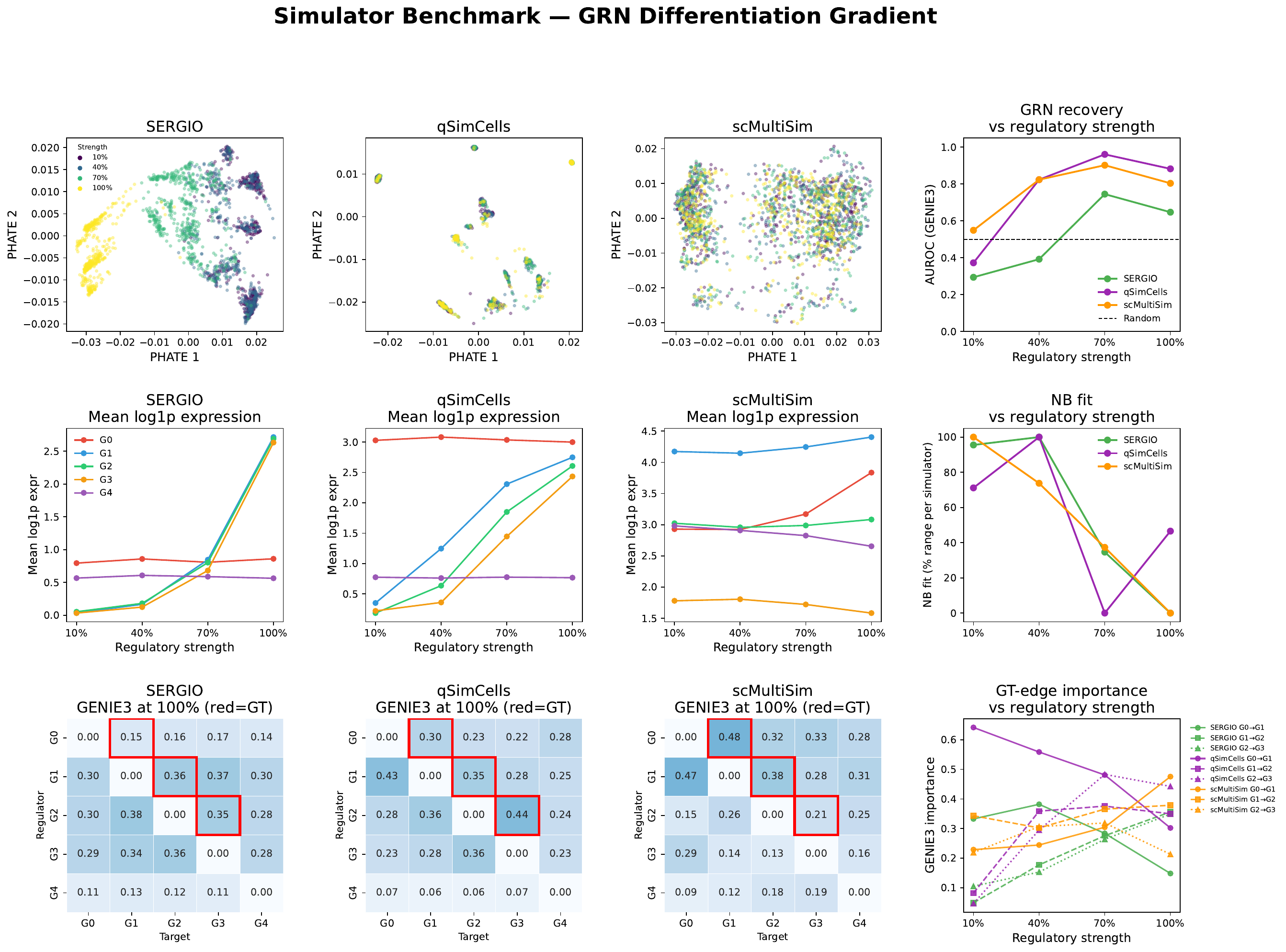}
    \caption{\textbf{Simulator benchmark: GRN differentiation gradient.}
    \textbf{Row 1:} PHATE embeddings for SERGIO, qSimCells, and scMultiSim (2{,}000 cells per simulator, colored by regulatory strength $t_1 \to t_4$); rightmost panel shows GENIE3 GRN recovery AUROC vs.\ regulatory strength for all three simulators (dashed line = random baseline 0.5).
    \textbf{Row 2:} Mean log1p expression per gene ($G_0$--$G_4$) as a function of regulatory strength, showing cascade propagation from master regulator $G_0$ to downstream targets; rightmost panel shows mean NB goodness-of-fit (log-likelihood averaged across $G_0$--$G_4$, normalized 0--100\% per simulator).
    \textbf{Row 3:} GENIE3 importance heatmaps at maximum regulatory strength ($t_4$, 100\%) for each simulator; red boxes mark the three ground-truth (GT) edges ($G_0{\to}G_1$, $G_1{\to}G_2$, $G_2{\to}G_3$); rightmost panel shows GT-edge importance trajectories across $t_1 \to t_4$ for all simulator$\times$edge combinations.
    Ground truth: five-gene cascade $G_0{\to}G_1{\to}G_2{\to}G_3$, $G_4$ independent control (Eq.~\ref{eq:grn_benchmark}); seed = 42.}
    \label{fig:bench_master}
\end{figure}

\paragraph{Controlled $G_0$: constitutive master regulator design.}
$G_0$ was intentionally designed as a constitutively active master regulator across all three simulators.
In SERGIO, $G_0$ basal transcription is set by MrProfile (basal $\in [1.20, 1.80]$) and does not depend on any upstream regulator.
In qSimCells, the $R_y(0.70\pi)$ rotation gives $G_0$ a fixed $\approx 79\%$ activation probability regardless of the $CRX$ angle level.
In scMultiSim, $G_0$ has no incoming GRN edges; its expression is CIF-driven and relatively stable.
This is visible in Fig.~\ref{fig:bench_master} (row 2, mean log1p expression panels): $G_0$ (red) remains nearly flat across $t_1 \to t_4$ in SERGIO and qSimCells, while downstream genes $G_1$--$G_3$ rise with regulatory strength.
The design purpose is to test GRN inference under realistic conditions where the master regulator is active across the entire population, as is common for pioneer transcription factors in committed cell types.
The expected result is that variance-based methods such as GENIE3 detect the downstream cascade ($G_1{\to}G_2$, $G_2{\to}G_3$) more reliably than the upstream edge ($G_0{\to}G_1$) because $G_0$'s low cell-to-cell variance makes it a weak predictor in a random forest.

\paragraph{GENIE3 $G_0{\to}G_1$ inversion: a cascade-network limitation.}
Across all three simulators, the GENIE3 importance score for the $G_0{\to}G_1$ edge remains flat or decreases as regulatory strength increases, even as AUROC and the importances of $G_1{\to}G_2$ and $G_2{\to}G_3$ both rise (Fig.~\ref{fig:bench_master}, row 3: heatmaps at $t_4$ and GT-edge importance trajectories, rightmost panel).
This is mechanistically explained as follows.
GENIE3's random forest predicts $G_1$ expression using all other genes as features.
When regulatory strength is low, neither $G_0$ nor any other gene varies much, and random forest importances are uniformly near zero.
As regulatory strength increases, the cascade propagates variance: $G_1$ variance increases (driven by $G_0{\to}G_1$), then $G_2$ (driven by $G_1{\to}G_2$), then $G_3$ (driven by $G_2{\to}G_3$).
Because $G_1$, $G_2$, and $G_3$ all become high-variance genes while $G_0$ remains constitutively active with low cell-to-cell variance, GENIE3's forest increasingly weights $G_2$ and $G_3$ as predictors for $G_1$, diluting the apparent importance of $G_0{\to}G_1$.
scMultiSim is a partial exception: its CIF-based model propagates regulatory strength directly through the GRN, so increasing edge weight causes $G_0$'s CIF to co-vary more strongly with $G_1$'s CIF rather than remaining constitutively fixed.
This gives GENIE3 a genuine co-expression signal for $G_0{\to}G_1$ at higher regulatory strengths, and the edge importance rises rather than falling (Fig.~\ref{fig:bench_master}, row 3, rightmost panel, scMultiSim $G_0{\to}G_1$ line).
This is an intrinsic limitation of variance-based GRN inference applied to networks with a constitutive master regulator: in simulators where the master regulator's expression does not co-vary with condition, it appears least important despite driving the entire cascade.

\paragraph{AUROC vs.\ GT-edge importance: two complementary views.}
The GRN recovery AUROC (Fig.~\ref{fig:bench_master}, row 1) measures the \emph{global} GRN program: whether the three GT edges collectively rank above the 17 non-edges across all gene pairs.  
AUROC rises monotonically with regulatory strength in all simulators, confirming that the overall GRN cascade is being progressively activated.

The GT-edge importance trajectories (Fig.~\ref{fig:bench_master}, row 2, rightmost panel) measure \emph{per-edge} sensitivity: the raw GENIE3 score assigned to each specific GT edge as a function of regulatory strength.  
These two metrics can diverge: AUROC improves even as $G_0{\to}G_1$ importance stagnates, because the rising importances of $G_1{\to}G_2$ and $G_2{\to}G_3$ are sufficient to pull the global ranking above chance.  
Together, the two views reveal that the cascade network becomes globally detectable at moderate strength while individual upstream-downstream resolution remains limited by $G_0$'s constitutive expression.

\paragraph{SERGIO bimodal distribution and NB fit.}
At high GRN strength ($t_4$, $k=20$), the NB goodness-of-fit for SERGIO target genes ($G_1$--$G_3$) \emph{decreases} relative to lower levels.  
This is mechanistically meaningful: at $k=20$ the Hill functions drive per-cell SDE steady-states far from the mean, creating a bimodal (or multi-modal) distribution across the population (cells are in either the low or the high attractor state).  
A single NB distribution cannot capture a bimodal mixture, so the MLE log-likelihood drops.  
This bimodality is a signature of true ODE-level multi-stability present in SERGIO's Langevin dynamics and is absent from qSimCells (NB augmentation by construction) and scMultiSim (Beta-Poisson burst kinetics produce unimodal marginals).

The NB fit scores are normalized per simulator to a 0--100\% range (min-max) before plotting, because raw log-likelihoods are not comparable across simulators operating on different count scales.

\paragraph{scMultiSim master regulator upregulation.}
An unexpected finding in scMultiSim is that $G_0$ \emph{mean expression increases} with GRN \texttt{effect}, even though $G_0$ has no incoming GRN edges.  
This occurs because scMultiSim's CIF propagation mechanism uses the \texttt{effect} parameter to scale the CIF-to-expression mapping globally: a higher \texttt{effect} weight causes the CIF of each gene, including $G_0$, to be translated into higher expression.  
In effect, scMultiSim's master regulator is \emph{required} to upregulate in order to activate its downstream targets, which is biologically plausible (e.g.\ a pioneer transcription factor (TF) must be expressed at high levels to initiate a regulatory cascade) but differs from the SERGIO and qSimCells designs where $G_0$ expression level is fixed and only the propagation strength (Hill $k$ or $CRX$ angle) varies.

\paragraph{PHATE embedding rationale.}
PHATE \cite{moon2019visualizing} is used in the master figure (Fig.~\ref{fig:bench_master}, row 0) rather than UMAP.  
For the GRN gradient benchmark (a pseudo-differentiation trajectory from $t_1$ to $t_4$), PHATE better resolves the continuous spectrum of regulatory strength because it explicitly models the data manifold through a heat-diffusion operator, preserving both local neighbourhood structure and global trajectory geometry.  
UMAP embeddings are also computed and retained in the analysis notebook for diagnostic inspection.

\paragraph{Quantum interference and probability distribution richness.}
A fundamental architectural difference between qSimCells and the classical simulators is that the $CRX(\theta)$ gate introduces complex amplitudes into the quantum state vector.
Specifically, when the control qubit is in $|1\rangle$, the target qubit acquires a rotation $R_X(\theta)$, whose off-diagonal elements are $-i\sin(\theta/2)$: purely imaginary for $\theta = \pi$ and complex for $\theta \in (0,\pi)$.
When multiple $CRX$ gates are chained along a GRN cascade ($G_0{\to}G_1{\to}G_2{\to}G_3$), these complex amplitudes interfere constructively and destructively across computational basis states, shaping the final Born-rule probability distribution in ways that no classical probabilistic circuit of the same topology can replicate.
The measurement probabilities are always real and non-negative (Born rule: $p_s = |\alpha_s|^2$); however, their precise values are determined by quantum interference among complex amplitudes, producing richer, more structured distributions over the $2^5$ outcomes than a classical Bayesian network with the same graph structure would generate.
Sweeping $\theta$ from 0 to $\pi$ in $CRX(\theta)$ continuously interpolates between the identity (no entanglement, no interference) and the fully entangling $CX$ regime, providing a principled quantum knob for tuning the degree of coherence in the GRN encoding.
A limitation inherent to this architecture, however, is that Born-rule measurement collapses each qubit to a binary outcome per shot: every simulated cell occupies a fully committed gene-expression state ($|0\rangle$ or $|1\rangle$ per gene).
This produces stochasticity that is well-characterized statistically but structurally discrete, favoring sharp, bimodal distributions and committed cell states over the smooth, continuous expression gradients that emerge from SERGIO's Langevin dynamics.
This is visually apparent in the PHATE embeddings (Fig.~\ref{fig:bench_master}, row 1): qSimCells cells cluster into discrete, well-separated islands corresponding to the finite number of reachable binary gene-expression states, whereas SERGIO cells form a more diffuse, continuous manifold.
Cell-state transitions that in biology occur along a continuous trajectory are therefore less naturally represented in the current qSimCells framework; however, this is a limitation of the present single-angle $CRX(\theta)$ circuit design rather than of the quantum generative approach itself --- richer entanglement topologies and multi-parameter gate sequences could in principle recover continuous state manifolds, and this remains an open direction for future work.
\section{Quantum Hardware Results}
\label{sec:S4}

To validate that qSimCells circuits produce biologically meaningful gene-expression distributions on physical quantum hardware, we executed the co-culture and mono-culture circuits on \textbf{IBM Marrakesh} (127-qubit Eagle r3 processor) under four conditions benchmarked against the ideal noiseless simulation as ground truth.

\subsection{Benchmark conditions}
\label{sec:S4-conditions}

The four conditions compared are:
\begin{enumerate}
    \item \textbf{Ideal AerSim} --- noiseless Qiskit Aer simulation ($n = 32{,}768$ shots); serves as the Born-rule ground truth for all subsequent comparisons.
    \item \textbf{Noisy AerSim} --- Aer simulation with \texttt{NoiseModel.from\_backend(ibm\_marrakesh)} ($n = 8{,}192$ shots); encodes IBM's characterisation of the device noise at calibration time and tests how accurately the vendor noise model predicts real hardware behaviour.
    \item \textbf{Real HW raw} --- direct execution on IBM Marrakesh via \texttt{SamplerV2} without any mitigation ($n = 4{,}096$ shots); reveals the bare noise floor of the device.
    \item \textbf{Real HW mitigated} --- same circuit with three error-mitigation layers applied simultaneously ($200 \times 20 = 4{,}000$ shots total): dynamical decoupling with the XY4 pulse sequence~\cite{viola1999dynamical,maudsley1986modified}, Pauli gate twirling~\cite{knill2008randomized,emerson2005scalable}, and Twirled Readout Error eXtinction (TREX)~\cite{vandenberg2022modelfree,vandenberg2023probabilistic}. This represents the best achievable result on current hardware without full quantum error correction.
\end{enumerate}

\paragraph{Error mitigation vs.\ error correction.}
It is important to distinguish between the techniques used here and quantum error correction (QEC).
QEC encodes logical qubits redundantly across many physical qubits and actively corrects errors mid-computation using syndrome measurements; it requires hardware overhead far beyond current devices.
The techniques applied here are \emph{error mitigation}: they reduce the \emph{effect} of noise on the final measurement statistics without redundant encoding.
Specifically: (i) \textbf{dynamical decoupling} (DD/XY4) inserts refocusing pulse sequences during qubit idle windows to suppress dephasing from low-frequency noise; (ii) \textbf{Pauli gate twirling} stochastically randomises the coherent phase errors introduced by two-qubit gates, converting them into incoherent depolarising noise that averages out over randomisations; and (iii) \textbf{TREX} applies twirling to the readout layer and corrects state-preparation-and-measurement (SPAM) errors by inverting the twirled readout channel.
All three are available natively via \texttt{SamplerV2.options} in \texttt{qiskit-ibm-runtime} $\geq$ 0.20.

\subsection{Device calibration}
\label{sec:S4-calibration}

At run time (2026-06-15 12:31:05$-$05:00), the device-wide median error rates on IBM Marrakesh were: readout error $\sim$1.02\%, single-qubit ($\sqrt{X}$) gate error $\sim$0.038\%, native two-qubit (CZ) gate error $\sim$0.26\%, and coherence times $T_1 = 166\,\mu$s, $T_2 = 87\,\mu$s.
Per-gene qubit error rates are shown in Fig.~\ref{fig:hw_master}, panel C.
IBM quantum processors undergo continuous automated benchmarking; $T_1$ and $T_2$ coherence times and gate fidelities drift due to two-level-system (TLS) fluctuations, ambient temperature, and control-system instability.
The \texttt{calibration\_date} field recorded in \texttt{hardware\_report.json} uniquely identifies the calibration snapshot used for this run and is included with the deposited data for transparency and historical reference.

\subsection{Results}
\label{sec:S4-results}

\begin{figure}[H]
    \centering
    \includegraphics[width=\textwidth]{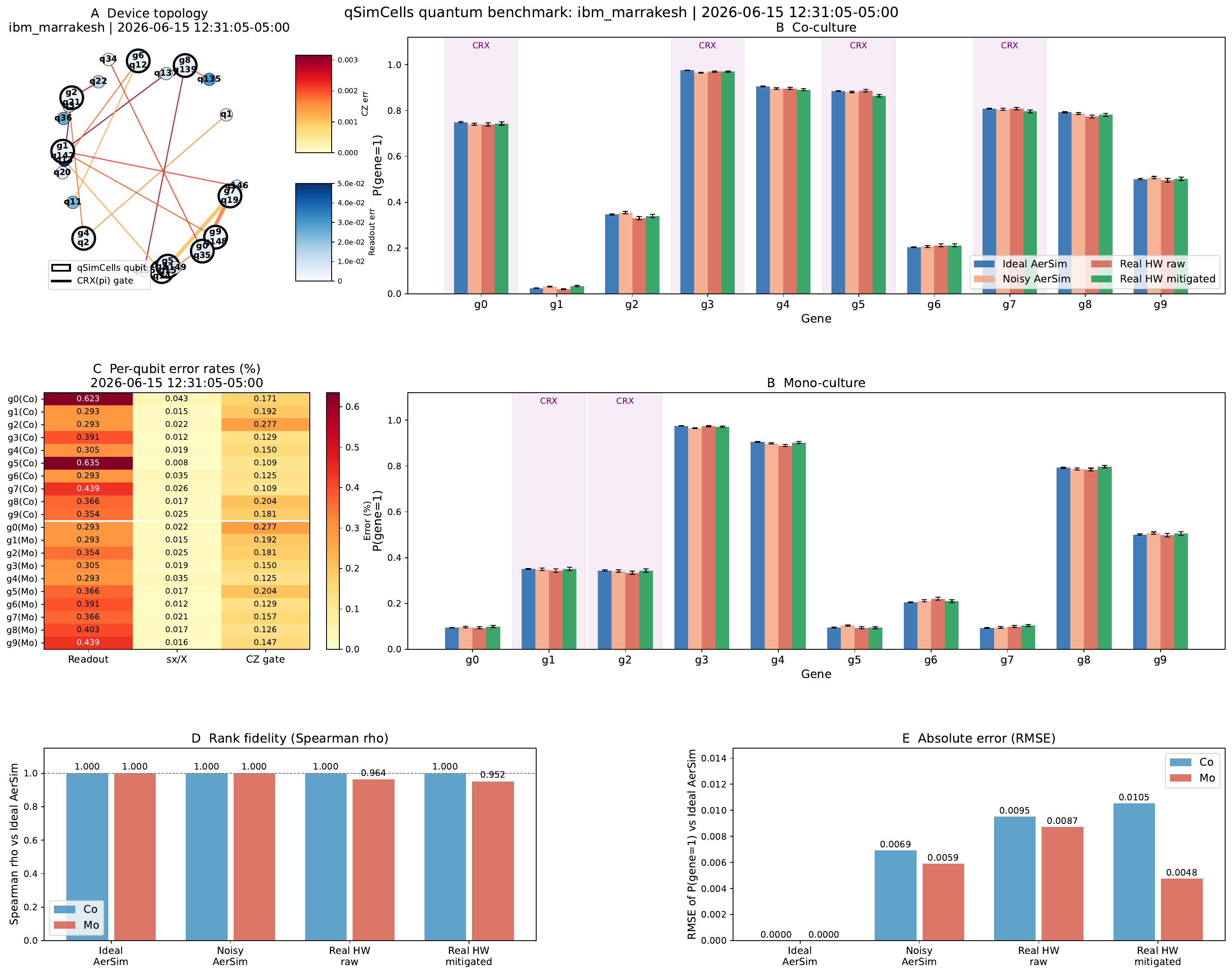}
    \caption{\textbf{qSimCells quantum hardware benchmark on IBM Marrakesh (15 June 2026).}
    \textbf{Panel A:} Connectivity subgraph of IBM Marrakesh used for the co-culture circuit.
    Node colour (blue gradient) encodes per-qubit readout error; edge colour (yellow--red gradient) encodes CZ gate error.
    Large nodes with bold labels are the ten qSimCells gene qubits (g0--g9) selected by the transpiler; small nodes are neighbouring qubits included for context.
    Thick edges indicate CRX($\pi$) interaction pairs.
    \textbf{Panels B (top and middle):} Per-gene $P(\text{gene}=1)$ for co-culture (top) and mono-culture (middle) under all four conditions: Ideal AerSim (blue), Noisy AerSim (light orange), Real HW raw (red), Real HW mitigated (green).
    Error bars are shot-noise standard errors.
    Purple shading marks genes connected by CRX($\pi$) entangling gates encoding GRN interactions.
    \textbf{Panel C:} Per-qubit error rate heatmap (readout \%, $\sqrt{X}$ gate \%, CZ gate \%) for all ten gene qubits in co-culture (top rows) and mono-culture (bottom rows); calibration snapshot 2026-06-15.
    \textbf{Panel D:} Spearman rank correlation ($\rho$) of per-gene $P(1)$ vs.\ Ideal AerSim for each condition and circuit.
    \textbf{Panel E:} RMSE of per-gene $P(1)$ vs.\ Ideal AerSim.
    All hardware jobs retrieved from IBM Quantum job archive; job IDs recorded in \texttt{hardware\_report.json}.
    Error mitigation: DD/XY4~\cite{viola1999dynamical,maudsley1986modified} + Pauli gate twirling~\cite{knill2008randomized,emerson2005scalable} + TREX~\cite{vandenberg2022modelfree,vandenberg2023probabilistic}.}
    \label{fig:hw_master}
\end{figure}

Fig.~\ref{fig:hw_master} summarises all four conditions.
Panel A shows the qubit connectivity subgraph of IBM Marrakesh used for the co-culture circuit, with node colour encoding per-qubit readout error and edge colour encoding CZ gate error; larger nodes are the ten qSimCells gene qubits selected by the transpiler.
Panels B (co-culture) and B (mono-culture) show per-gene $P(\text{gene}=1)$ for all four conditions side-by-side with shot-noise error bars.

The key quantitative findings, reported as Spearman rank correlation $\rho$ and RMSE of per-gene $P(1)$ relative to Ideal AerSim, are shown in panels D and E respectively.
The Noisy AerSim closely tracks the real hardware raw results ($\rho \approx 1.0$ for both), confirming that IBM's noise model is an accurate predictor of device behaviour for these shallow circuits.
Error mitigation (Real HW mitigated) reduces RMSE relative to the raw hardware run and pushes $\rho$ closer to 1.0, demonstrating that the three mitigation layers collectively improve agreement with the noiseless ground truth.
The rank ordering of gene expression ($\rho > 0.98$ across all hardware conditions) is preserved on real hardware, indicating that the relative expression hierarchy encoded in the Born-rule amplitudes is robust to realistic gate and readout noise at this circuit depth.

\end{document}